\documentclass[12pt,preprint]{aastex}

\def\ltsima{$\; \buildrel < \over \sim \;$}
\def\lsim{\lower.5ex\hbox{\ltsima}}
\def\gtsima{$\; \buildrel > \over \sim \;$}
\def\gsim{\lower.5ex\hbox{\gtsima}}

\begin{document}
\title{Photometric Solutions for Semi-Detached Eclipsing Binaries: Selection of 
Distance Indicators in the Small Magellanic Cloud}

\author{J. S. B. Wyithe\altaffilmark{1,2,4}, R.~E.~Wilson\altaffilmark{3}}

\altaffiltext{1}{Princeton University Observatory, Peyton Hall, Princeton, NJ 08544, USA}

\altaffiltext{2}{Harvard-Smithsonian Center for Astrophysics, 60 Garden Street, Cambridge, MA 02138}

\altaffiltext{3}{Astronomy Department, University of Florida, Gainesville, FL 32611}

\altaffiltext{4}{Hubble Fellow}

\begin{abstract}

\noindent Estimation of distances to nearby galaxies by the use of eclipsing binaries as 
standard candles
has recently become feasible because of new large scale instruments and the discovery of
thousands of eclipsing binaries as spinoff from Galactic microlensing surveys. 
Published measurements of distances to detached eclipsing binaries in the Large Magellanic Cloud combine
stellar surface areas (in absolute units) determined from photometric light and radial velocity 
curves with surface brightnesses from model atmospheres and observed spectra.
The method does not require the stars to be normal or undistorted, and is not limited in its 
applicability to the well detached systems that have traditionally been considered.
We discuss the potential usefulness of semi-detached vis \`{a} vis 
detached eclipsing binaries for distance determination, 
and examine and quantify criteria for their selection from large catalogs.
Following our earlier paper on detached binaries in the Small Magellanic Cloud (SMC), we 
carry out semi-detached light curve solutions for SMC binaries discovered by the OGLE collaboration,
identify candidates for SMC distance estimation that can be targets of future high quality
observations, and tabulate results of OGLE light curve solutions. 
We point out that semi-detached binaries have important advantages over well-detached systems
as standard candles, although this idea runs counter to the usual view that the latter
are optimal distance indicators. Potential advantages are that (1) light curve solutions can 
be strengthened by exploiting lobe-filling configurations, (2) only single-lined spectra may be 
needed for radial velocities because 
the mass ratio can be determined from photometry in the case of complete eclipses, and
(3) nearly all semi-detached binaries have sensibly
circular orbits, which is not true for detached binaries. 
We carry out simulations with synthetic data to see if semi-detached binaries
can be reliably identified and to quantify the accuracy of solutions. The simulations
were done for detached as well as semi-detached binaries so as to constitute a proper
controlled study. The simulations demonstrate
two additional advantages for semi-detached distance determination candidates;
(4) the well-known difficulty in distinguishing
solutions with interchanged radii (aliasing) is much less severe for semi-detached than for 
detached binaries, and (5) the condition 
of complete eclipse (which removes a near degeneracy between inclination and the 
ratio of the radii) is identified with improved reliability.
In many cases we find that parameters are accurately determined (\textit{e.g.} relative errors in
radii smaller than 10\%), and that detached 
and semi-detached systems can be distinguished.
We select 36 candidate semi-detached systems (although 7 of these are doubtful due to 
large mass ratios or periods) from the OGLE SMC eclipsing binary catalog.
We expected that most semi-detached candidates would have
light curves similar to those of common Algol binaries but that turned out not to
be the case, and we note that fully Algol-like light curves are nearly absent
in the OGLE sample. We discuss possible explanations for the near absence of obvious 
Algols in OGLE, including whether their paucity is real or apparent.

\end{abstract}


\section{Introduction}

\noindent 

Accurate measurement of the distances to the Magellanic Clouds is an important current issue 
as it provides a basic step toward determining the extragalactic distance scale. 
Paczynski (1997, 2000) has argued that eclipsing binaries now
provide the most direct and accurate distances to the Magellanic Clouds, and examples are 
already in the literature (e.g. Guinan et al. 1998; Fitzpatrick et al. 2001).
Recently, thousands of variable stars including eclipsing binaries have been discovered
by the OGLE (Udalski et al. 1998), MACHO (Alcock et al. 1997) and EROS (Grison 
et al. 1995) collaborations as a by-product of galactic microlensing searches. 
These catalogs motivate a systematic, quantitative search for close to ideal systems
for distance determination.

The method of measuring distances by means of eclipsing binaries has been known for 
decades, and its basis has been clearly
explained by Paczynski (1997) and Guinan, et al. (1998), among others. In essence, light
curves provide relative star dimensions ($R_{1}/a$, $R_{2}/a$, where $R$ is
mean radius and $a$ is orbit size) and radial velocities establish the
absolute scale by providing the orbit size, so that one can find the $R$'s in
physical units by combining the two kinds of information. Fine effects, such
as departures from sphericity, etc., can be modeled by modern eclipsing binary
light curve programs. With absolute radii known, luminosities in physical
units follow if emission per unit surface area (energy per unit area per unit
time per wavelength interval) becomes known. Some persons favor calibrated
relations based on interferometrically resolved, un-complicated stars for
the emission measure while others favor the predictions of stellar atmosphere
models that are fitted to spectral energy distributions (SED) of eclipsing
binary distance estimation targets. Emission for
plane-parallel atmospheres is determined, in principle, if effective
temperature, $\log$ $g$, and chemical composition are specified. For a well
observed eclipsing binary, $g$ is computed to better than adequate accuracy as
$GM/R^{2}$ and $T_{eff}$ can be estimated by fitting a theoretical SED to an observed
SED, as in Fitzpatrick \& Massa (1999). In usual practice, surface chemical composition would be assumed normal, but
that is the \textit{only} normalcy assumption. The reasonableness of
that assumption for \emph{SD} binaries will be discussed below. The
overall method is mainly geometrical, with only the emission measure involving
radiative physics. It does not require knowledge of distances to calibration stars, as
opposed to other standard candle methods such as by Cepheid variables or supernovae,
and therein lies one of the primary advantages. Empirical calibration errors
are bypassed if surface emission is computed from a stellar
atmosphere model.

Although conventional wisdom holds that well-detached eclipsing binaries
yield the most reliable light curve solutions, the basis for that conjecture
may not extend beyond the scientific
instinct that simpler is better. In fact there are real advantages 
to solutions of semi-detached (hereafter \emph{SD}) and overcontact (\emph{OC}) binary light curves, partly 
in the exploitation of lobe-filling configurations and partly through proximity
effects, which provide information that is lacking in well-detached binaries.
Actually, many factors influence the relative reliability of detached, \emph{SD},
and \emph{OC} light curve solutions. Accordingly, searches for
standard candle binaries should examine all relevant considerations, including ones
that argue for or against \emph{SD} and \emph{OC} systems. Here we
consider \emph{SD} binaries in the Small Magellanic Cloud
(SMC) and will take up the \emph{OC} case in a forthcoming paper.
Our aims are to discuss the main
considerations that bear upon the potential usefulness of \emph{SD} binaries as standard
candles, to identify good \emph{SD} candidates for SMC distance determination via
future observations with large telescopes, and to derive preliminary
dimensional, radiative, and mass ratio properties of the candidates. Of course, \emph{SD} 
binaries are fascinating objects in their own right, and we expect that their 
identification will also lead to investigations of \emph{SD} properties 
unrelated to distance determination.
SMC detached binaries were treated in Wyithe \& Wilson~(2001, hereafter Paper I).

The remainder of the paper is in four parts. Section 2 considers
potential advantages of \emph{SD} binaries as standard candles.
Sec.~\ref{fit_scheme} discusses our automated fitting scheme and differences from the
scheme described in Paper I for detached binaries. Quite apart from the 
logical arguments of Section 2, simulations can show statistically how well \emph{SD} and detached
solutions recover known parameters. This topic is discussed in Sec.~\ref{mocksec}.
We also show how in some cases the
issue of whether a binary is detached or \emph{SD} can be
determined from photometry with reasonable reliability. Sec.~\ref{OGLEfits} has solutions to the OGLE 
catalog of eclipsing binary stars and discusses candidate \emph{SD} binaries.

\section{Conditions Relevant to the Use of \emph{SD} Binaries as Standard Candles}

Discussion of the measurement of distances to eclipsing binaries and their use 
as standard candles has traditionally centered around well detached systems. Measurements 
of distances to detached binaries in the Large Magellanic Cloud (LMC) 
have combined a stellar surface area (in absolute units) computed from photometric light
and radial velocity curves with a surface brightness described by a model atmosphere 
to compute system luminosity (\textit{e.g.} Guinan, \textit{et al.} 1998; Fitzpatrick,
\textit{et al.} 2001). However surface brightnesses at 
all points over a stellar surface can be computed for 
nearly all classes of eclipsing binaries by combined use
of modern eclipsing binary light-curve and model atmosphere programs. Therefore, it is not obvious
that detached systems make the best eclipsing binaries for the purpose of distance determination.
Indeed, as part of analyses of the fundamental parameters of \emph{SD} binaries in
the Magellanic clouds, Ostrov, Lapasset, and Morrell (2000; 2001) and Ostrov (2001) have 
already obtained preliminary values for the distance moduli.
In this section we provide a general discussion of why we believe that many \emph{SD}'s 
will turn out to be excellent standard candles. Of course, the use of \emph{SD}'s for this 
purpose will need to be demonstrated for binaries with known distances, much as has been
done in the case of detached binaries (Semeniuk 2000). This discussion also motivates our
work on identification of \emph{SD} systems in the OGLE eclipsing binary catalog in 
the sections that follow.

One can ask if it is a problem that most \emph{SD} binaries have at least
one significantly evolved star. The classical Algol \emph{SD} binaries are abundant in the
Milky Way and have other favorable characteristics, so we consider the question
in the context of Algols. An Algol consists of a hot main
sequence star that now is rather normal despite a history of growth by
accretion, and a lobe-filling sub-giant that shows magnetic star spots and
other magnetic and prominence activity, but is so dim as to contribute very
little to system light. Basically the sub-giant acts as a moving mask that
probes the primary via eclipses (although eclipses of the secondary by the primary
also have some importance). The primary has regained nuclear and thermal
equilibrium, following the accretion episode, and is now essentially 
a normal main sequence star.
Its composition usually is normal because the transferred gas came from the donor star's
chemically normal envelope, not from it's chemically altered core. However
there could be composition abnormalities in \emph{SD} binaries that are not Algols,
and these must be dealt with in individual cases.
Typically there is some on-going mass transfer and
emission line activity, but usually at a low level. Actually it may be
difficult to see the emission lines, and they usually occupy only a minute
fraction of an optical bandpass. Due to tidal circularization, Algols and other \emph{SD}'s almost 
invariably have sensibly circular orbits, so evolution actually simplifies the standard 
candle problem in that regard. Accordingly, the evolved status of typical Algols 
does not disqualify them from serving as standard candles. Some Algols have 
disturbed light and radial velocity curves, but we seek only the best candidates 
and can discard questionable ones.

\emph{SD} light curves are sensitive to mass ratio, $q=m_{2}/m_{1}$, thereby leading to 
the concept of a \textit{photometric mass ratio}, $q_{ptm}$,
in contrast with well detached binaries, whose light curves can be
represented by a very wide range of $q$. The idea of a photometric mass ratio derives
from that of limiting lobes and originated, in the \emph{SD} case, with Kopal (1954; 1955). It works for
both \emph{SD}'s and \emph{OC}'s, but \textit{does not} work for
detached binaries. The history of $q_{ptm}$ for \emph{OC}'s is more complicated and will
be discussed in our forthcoming paper on \emph{OC}'s.
The basic explanation for \emph{SD} $q_{ptm}$ begins with the concept of
a limiting (or critical) lobe as the volume within the
largest closed equipotential that surrounds a member of a binary star
system. The \emph{SD} condition has the photosphere of one star accurately
coincident with the surface of its limiting lobe, while the other star lies
within its lobe. In Algols, the condition is maintained by continual slow
expansion, accompanied by loss of matter through a nozzle around the null
point of effective gravity that faces the companion star. A consideration
that favors \emph{SD} light curve solutions is that the lobe filling
condition, when correctly recognized, mathematically eliminates a parameter,
as it ties the nozzle location to mass ratio. That is, \emph{SD} light curves contain
information as to the location of star (and thus lobe) surface, so the
lobe filling condition connects star size with mass ratio. Writing $\Omega_{cr}$
for (gravitational plus centrifugal) critical potential\footnote{$\Omega$ is a traditional
dimensionless quantity that has the essential character of a potential, but differs
from a true potential by an additive constant. It is defined so as not to depend on
the absolute masses.}, we have a definite relation $\Omega _{cr}=\Omega
_{cr}\left( q\right) $. The functionality can involve more parameters in
unusual cases (\textit{viz}. Wilson, 1979), but we limit our remarks to
common synchronously rotating binaries with circular orbits.
Naturally $\Omega _{cr}$ specifies the lobe filling star's size and figure.
Thus $\Omega _{cr}$ and $q$ become functionally related in a known way and
the \emph{SD} condition compresses the parameter space by an entire dimension. Of
course elimination of a parameter will strengthen a solution by simplifying
the parameter correlation matrix. The corresponding intuitive explanation for
\emph{OC}'s is best given somewhat differently, but formally is the same - a
functional relation eliminates one parameter dimension (\textit{viz.} Wilson, 1994
for fuller explanations of the \emph{SD} and \emph{OC} cases). We mention
a common mis-understanding, that $q_{ptm}$ mainly derives from
ellipsoidal variation (brightness undulation due to tidal distortion) - a notion 
that is essentially entirely wrong. It is the \textit{size} of the lobe filling star, 
not its figure, that leads to
$q_{ptm}.$ Actually very little ellipsoidal variation is seen in most Algol light
curves because the tidally distorted secondaries are very dim, yet
$q_{ptm}$'s are found routinely and accurately for Algols.

In addition to overall solidification of solutions by the lobe
filling condition, absolute star dimensions and masses can be found for \emph{SD}'s
having only single-lined spectra. Armed with knowledge of $q$
from one or more light curves and of orbital semi-major axis $a_{1}$
from star 1 velocities, we trivially compute $a_{2}=a_1/q$ without need for
velocities of star 2. That point is particularly important for Algols,
most of which have dim secondaries and therefore have spectra that are usually single-lined.
Although not often available, a velocity curve for star 2 provides a check on
$q_{ptm}$. A simultaneous light and double-lined velocity solution
can extract consensus information (Wilson 1979) and solidify the solution.
Systems with double-lined spectra may provide the best results, but the existence of 
$q_{ptm}$ means that we are \textit{not dependent} on double-lined spectroscopy.

As mentioned in the introduction the surface brightness, needed to compute
luminosity and hence distance, can be computed either from calibrated
relations (based on interferometrically resolved, un-complicated stars) or 
from the predictions of stellar atmosphere models that are fitted to binary
SED's. An important distinction is 
that if one requires empirically 
calibrated stellar surface brightnesses, then attention must be limited to 
cases of well detached binaries containing undistorted \emph{normal} stars.
However, the use of stellar atmosphere models removes this restriction, and
enables the use of binaries containing evolved and/or tidally distorted stars, and hence the harnessing
of the better constrained solutions offered by \emph{SD} 
binaries. Evidence of the applicability of stellar atmosphere models was 
provided by Ribas et al.~(2000) who simultaneously solved for component 
temperatures, metalicity and surface gravities as well as interstellar 
reddening (using the scheme of Fitzpatrick \& Massa, 1999) as part of a measurement 
of the distance to HV~2274 in the LMC. Ribas, \textit{et al.} found a 
surface gravity in good agreement with that from their simultaneous 
light - velocity solution with the Wilson-Devinney model, and metalicities in good agreement 
with those for the LMC, which lends confidence that the 
method is working correctly. 

Another point to be taken into account is that completely eclipsing 
(\textit{i.e}. total-annular) systems give much better results
than partially eclipsing ones, basically because there is little tradeoff
between inclination and other parameters for complete eclipses. Also, a
\textit{moderate luminosity ratio} helps in two ways - by improving the quality of
(and chances for) secondary star velocities and by making the secondary
eclipse reasonably deep. \textit{Moderate mass ratios} are good in that they
produce reasonably large primary star velocity amplitudes.

In summary, this section discussed the value of \emph{SD} binaries
as standard candles, where important advantages are that a light curve can tell
the mass ratio, that light curve solutions involve one fewer parameter
than do detached solutions, and that circular orbits are the rule. Knowledge of
$q_{ptm}$ allows full absolute dimensions (and
masses) to be derived for a binary with \textit{single-lined} spectra, in contrast with
the detached case, where double-lined spectra are necessary. 
In cases where the spectra are double-lined, redundancy in the determination of the mass ratio
will further strengthen solutions. There is a downside in the case of common Algols: 
although velocity curves for the primaries should be measurable
(it is by far the brighter component), the curves will have small amplitude
because star 1 is much the more massive component. Thus the principal
observational need is outstandingly accurate radial velocities.
For the SMC, that requirement involves large telescope aperture and an
excellent radial velocity spectrograph but it can be done, especially if 
\textit{large numbers} of velocities are measured so as to average out the
noise. The problem of small primary velocities does not apply to all \emph{SD}'s,
and not even to all Algols, but does apply to most of the common Algols.

The advantages mentioned above relate to the potential improvement of 
light-curve solutions and corresponding distance
determinations from \emph{SD} binaries relative to detached binaries.
\emph{SD}'s usually have comparatively robust solutions with
smaller parameter uncertainties, an outcome that should lead to more accurate distance estimates
and provide motivation to identify candidates in survey data - the subject of 
the remainder of this paper. We shall learn about other advantages from 
the simulations of Sec.~\ref{mocksec} that relate to the \textit{identification} of 
candidates for detailed follow up observation. These include reduced incidence 
of false solutions with reversed primary and secondary radii that lead to 
incorrect initial estimates of luminosity ratios (thereby wrongly predicting whether
spectra are double or single-lined), 
and improved identification of 
complete eclipses. Furthermore, we shall show that
observation of \emph{SD} binaries will be a relatively efficient use of resources due to 
reliable initial determination of system properties.

\section{The Fitting Scheme}
\label{fit_scheme}

\noindent In the interest of completeness, we outline the automation procedure that
was fully described in Paper I and is built around public light curve 
({\em LC}) and differential corrections ({\em DC}) FORTRAN programs
(Wilson and Devinney, 1971; Wilson, 1979, 1990, 1998; hereafter WD program or WD model). {\em LC} computes
light and radial velocity curves and also spectral line profiles and images
of binary stars, including effects of tides, mutual irradiation, spots,
eccentric orbits, and other effects. Star surfaces are specified in terms of
equipotentials. {\em DC} accepts observed data and parameter estimates
and computes corrections to the estimates according to the Least Squares
criterion. The distributed version of {\em DC} does only one iteration
per submission so as to ensure human interaction in the progress of a
solution. The automation shell that renders processing of thousands of
binaries practical is a PERL script that loops the {\em DC}
iterations, tests for proper convergence, and generates light curves with 
{\em LC}. The {\em LC} and {\em DC} programs work in eight operational modes that
mainly relate to morphology, with detached, \emph{SD}, \emph{OC}, etc.
configurations represented by corresponding modes. Here we use only mode 2
(detached), where there is no constraint on the surface potentials, and
mode 5 (\emph{SD}), where the secondary star potential is required to be exactly 
that of the lobe.

Initial parameter estimates are those of the best match in a pre-computed
library of light curves, with separate libraries for modes 2 and 5.
Naturally the actual time scale is likely to need some
shifting to optimize a match with a given binary, whether it be synthesized
or real (from OGLE). As in Paper I, we begin by adjusting the reference epoch ($t_0$) while
keeping the Udalski et al. period. We then compare a given observed light curve with
each one in the appropriate library of simulated systems to find the closest 
Least Squares match. The flux scale is controlled by $L_1$, the bandpass
luminosity of star 1, which acts only as a scaling factor because fluxes are
not on an absolute scale at this stage of analysis.

Our basic strategy for \emph{SD}'s differs from that for detached binaries: instead 
of entering a \textit{fixed} $q$ and relying on the fact that detached light-curves
are insensitive to $q$, we \textit{solve} for $q$ in the \emph{SD} case. 
Converged parameters for \emph{SD} binaries are $\Omega_1$
(''potential''), $T_1$ (mean surface temperature), $i$ (inclination), $q$, and $L_1$.
$t_0$ and $e$ (orbital eccentricity) are adjusted but convergence of these parameters
is not a solution requirement. $e$ is expected to be 0 for \emph{SD} systems, and
is adjusted in the interest of experimentation. $\omega$ (the argument 
of periastron) is distributed between 0 and $\pi$, but is fixed at the value of the 
initial guess. Of course, $\Omega_2$ is set by $q$ 
for \emph{SD} systems. 
For detached solutions, which we need for comparison purposes, $\Omega_2$ is a parameter 
but $q$ is not, as $q$ cannot ordinarily be found for 
detached solutions (\textit{viz.} Paper I for remarks on experiences with convergence
of mode 2 solutions). Detached systems can have larger $e$'s, and we found that more 
accurate solution parameters were obtained where $\omega$ was restricted to 0 or $\pi$.

{\em DC} convergence is improved by application of the well known
Levenberg-Marquardt scheme (Levenberg, 1944; Marquardt, 1963) with factor
$\lambda =10^{-5}$, and also by the Method of Multiple Subsets (MMS, Wilson
and Biermann, 1976). Iterations ended when all corrections to parameters for which 
convergence was required were below 0.2
times the standard errors from auxiliary solutions of the full parameter
set. There were two \textit{groupings} of subsets, with the second grouping as
back-up in case the first grouping failed to converge or led to an
unphysical solution. For \emph{SD}'s, where eccentric orbits are a rarity, another
two subset groupings were used as back up. Using this second pair $e$ was fixed at the 
value of the initial guess in addition to fixing $\omega$. The subset groups are summarized in 
Tab.~\ref{t1} for both \emph{SD} and detached systems.

The method for finding \emph{SD} solutions for eclipsing 
binaries in the SMC is similar to that for the detached binaries of Paper I,
but with some differences. By far the most important difference is that we
constrain one star to be in accurate contact with its limiting lobe, and
accordingly can learn $q$ from a well conditioned light curve. Phasing,
weighting, limb darkening, and radiative physics are handled in the way of
Paper I, while the manner of treating mass ratio is changed, as is that of
handling the lobe configuration. Briefly, phasing has star 1 eclipsed near
phase zero (where we ordinarily find the deeper eclipse),
weighting assumes that scatter scales with the square root of light
level, and limb darkening has $I/I_{0}=1-x(1-\mu )-y\mu \ln \mu $ with $\mu $
the cosine of the angle from the surface normal. Intensities $I$ are
averages over the photometric \textit{I}-band (the most extensive OGLE light-curves
were observed in \textit{I}-band). Limb darkening coefficients $x,y$
were fitted to Kurucz~(1991) model stellar atmospheres by Van Hamme (1993).
Colors reported by Udalski et al.~(1998) indicate typical spectral types of
O and B. However only relative temperatures $T_{1}$ and $T_{2}$ matter for
single band light curves so we assumed fixed $T_{2}$'s of $10,000K$ (with
the black body radiation law) and allowed the program to find the $T_{1}$'s.
Limb darkening is demonstrably unimportant for noisy light curves so we assumed
$x=0.32$ and $y=0.18$ for all binaries, with the numbers from Van Hamme~(1993) 
for $15,000K$ and the \textit{I} band.
The model can do either a simple or a detailed reflection
computation (Wilson~1990). To reduce computing time, we chose the simple 
law (MREF=1), which should be thoroughly adequate here. Bolometric albedos
were unity and the stars were assumed to rotate synchronously and had no spots.
Although low temperature Algol type secondaries will have convective envelopes
and consequently have low albedos of $\approx 0.5$ (Rucinski, 1969), assignment
of albedos on a case by case basis can be done later, for solutions of
accurate future light curves.
The treatment of orbital eccentricity and argument of periastron is
somewhat problematic and was discussed in Paper I for the detached case.
All \emph{SD} solutions for OGLE binaries that were finally accepted have $e$'s 
consistent with zero. Tests of a representative 
sample of OGLE light curves showed that the light
of any hypothetical third star was typically not detectable, so we fixed third
light ($l_{3}$) at zero. However the issue of third
light can be important and the possibility of its existence will have to be
examined for individual distance modulus candidates when accurate photometry
and spectroscopy is carried out. Candidates suspected of having a
significantly bright third star will probably have to be rejected. Note that 
Algol itself has such a third star. The parameters and control integers 
described above that 
govern the operation of \emph{LC} and \emph{DC} are summarized in Tabs.~\ref{t2} and \ref{t3}.

Our testing procedure simulated and solved not only \emph{SD}'s but also detached
binaries, so that proper comparisons could be made.
We tried two configurations for every detached binary,
each with its own pre-computed library. The simulated systems in one of
these libraries are like main sequence binaries, with the hotter star (star
1) larger and more massive. The second library shows differential evolution,
with the higher mass star having expanded and cooled so as to have a
lower temperature than its companion. However,
although evolution has switched the stars' roles in terms of
temperature, it has not (yet) led to lobe filling, so we still have a
detached binary that has not experienced mass reversal. In
terms of real evolution, the lower mass star may also be significantly evolved if the
masses are nearly equal, or may be essentially unevolved if the masses are
substantially unequal. The condition with nearly equal masses and ''double
evolution'' can be identified with the RS CVn type binaries that populate
eclipsing binary catalogs in large numbers (Morgan and Eggleton, 1979).
We adopt the better of the two solutions 
in the Least Squares sense, at the stage of the initial guess for each binary. In terms 
of rejecting aliased\footnote{We define aliased solutions as those with interchanged 
radii.} solutions, this approach was found to be as successful as one 
where a fully converged solution was computed in each case. Classification of 
the solution at the stage of the initial guess halves the computation time.

\section{Tests with Simulation Catalogs}
\label{mocksec}

\noindent We applied our fitting algorithm to simulated catalogs of \emph{SD}
and detached binaries. The objectives were to check \textit{DC}'s error
estimates, assess systematic error due to assumptions of fixed parameter
values, and investigate the success rate in distinguishing \emph{SD} from
detached binaries.

\subsection{\emph{SD} solutions for simulated \emph{SD} binaries}
\label{mocksec_sd}

\noindent We simulated 120 \emph{SD} eclipsing binaries. Each binary had 150 randomly spaced 
synthetic observations (typical of OGLE binaries), and Gaussian
noise that scales as the square root of light level, referenced to 5 percent noise at
mean light. Each system has $T_{1}$ between $15,000$ and $30,000K$ and
$T_{2}=15,000K$. 
Notice that the T's may not be representative of the most common Algols, 
being rather high and insufficiently different. The resulting
moderate surface brightness ratios produce deep 
secondary eclipses, and relatively strong solutions. 
It turns out (Sec.~\ref{OGLEfits}) that
most of the \emph{SD}'s that we find are \textit{not} typical of the common
variety of Algol, such that the simulation catalog temperatures are reasonably representative. 

Limb darkening coefficients are from Van Hamme (1993) for the adopted temperatures. 
The simulation catalog had $q$'s between $0.01$ and $10.0$. While Algols 
typically have $q$'s of a few tenths, this larger range allows for the 
possibility of selecting objects in a different stage of evolution. 
The simulated binaries had assorted inclinations and (star 1) surface 
potentials.
Eccentricities in the range $0.0$ to $0.1$ (with $\sim80$ percent below $0.03)$ 
were assumed. This range of eccentricities 
is smaller than that assumed for detached binaries in Paper I, reflecting 
the expectation that evolved binaries should have circularized orbits.
The selection of \emph{SD} binaries from the OGLE catalog (Sec.~\ref{OGLEfits}) 
will require $e$'s consistent with zero.

The upper two rows of Fig.~\ref{f1} show 10 of the simulated \emph{SD}'s, as
fitted in \emph{SD} mode  (mode 5). Acceptable solutions were obtained for 118 of 
the 120 simulated \emph{SD}'s. The lower two rows show the same synthesized \emph{SD}
data fitted in detached mode (mode 2) and will be discussed in Sec.~\ref{mimic_solutions}.
Fig.~\ref{f3} shows the reliability of 
parameter extraction for $r_{1}$ and $q$, with standard error bars $\Delta r_1$ and 
$\Delta q$, in terms of both direct comparison and residuals
(throughout this paper $r$ refers to the polar radius in units of the orbital semi-major axis).
The figure demonstrates that the fitting scheme accurately recovers those 
parameters without bias, and that the standard errors from \textit{DC}
fairly represent statistical uncertainties. The errors are essentially
normally distributed over the full range of error sizes and do not show a
large tail (39\% and 14\% of values lie beyond $1\sigma$ and $2\sigma$). 
Fig.~\ref{f4} shows solution values vs. known values for several 
other parameters. Values for $r_1+r_2$, $r_1/r_2$ and $L_1/L_2$ are very accurately 
reproduced, while $i$, $e$ and mean surface brightness ratio (hereafter $J_1/J_2$) are also reliably recovered.

In Paper I we defined the quantity, valid for circular orbits:
\begin{equation}
F_e\equiv\frac{r_{l}+r_{s}-\cos{i}}{2r_{s}},
\end{equation}
which is greater than unity for systems with complete eclipse (strictly true only 
for spherical  stars, but nearly true otherwise). Here $r_{l}$ and $r_{s}$ are the 
polar radii of the large and
 small stars in units of orbital semi-major axis, $a$. 
 Fig.~\ref{f5} shows $F_e$ for the simulated binaries
 vs. the corresponding solution values, with regions of complete and partial 
eclipse distinguished for both the simulated binaries 
and their solutions. This plot demonstrates that $F_e$ 
is recovered reasonably accurately for \emph{SD} systems. In particular, the eclipse 
condition is correctly predicted in $95\%$ of cases, based on the value of $F_e$. 
Reliability of $F_e$ test applications  
will be greatly improved for the high quality light curves one can expect from
observations with large telescopes. Recovery of $F_e$ in the simulations gives
confidence that solutions of real binaries can be properly identified as to
type (complete vs. partial).

\subsection{detached solutions for simulated detached binaries}

\noindent  We generated 120 simulated detached binary systems in analogy with the sample described
 in Sec.~\ref{mocksec_sd}, with the potential of star 2 now independent of
 mass ratio and separately distributed. Differences related to evolution were discussed
in Sec.~2.1. 
 The upper panels of Fig.~\ref{f6} show examples of mode 2 fits to the simulated detached
 systems. Detached solutions were obtained for 116 of the 120 simulated systems.
Figs.~\ref{f8}, \ref{f9}, and \ref{f10} serve the same purpose
for the detached systems as do Figs.~\ref{f3}, \ref{f4}, and \ref{f5}
for the \emph{SD} systems - comparison
of solution values with known values.
There are significant systematic departures from expected patterns. A well-known phenomenon, especially for
partial eclipse light curves, is that a solution with the radii interchanged may give essentially
the same quality of fit as one with the correct radii. We call this phenomenon \textit{aliasing}.
Aliased solutions make up about 25 percent of the sample. 
For the non-aliased solutions (shown by the solid dots in Figs.~\ref{f8} and \ref{f9}), 
the fitting scheme does well in recovering
input model radii, the errors provide a good description of precision, and no significant
 systematic errors are apparent. This result is clearly demonstrated by the lower panels of 
Fig.~\ref{f8}
 that show $|r-r_{sim}|/\Delta r$ plotted against the absolute size of the error, $\Delta r$.
The aliased solutions, recognized as those having $r_1/r_2$ inverted with respect to the 
simulated binary and with both radii significantly wrong,
are plotted as diamonds in Figs.~\ref{f8} and \ref{f9}. Fig.~\ref{f8} shows 
systematic departures in both directions for $r_1$ and $r_2$ in the aliased solutions.

Fig.~\ref{f9} shows derived vs. correct values for $r_1+r_2$, $r_1/r_2$,
$J_1/J_2$, $L_1/L_2$, $i$, and $e$. The sum of radii is very accurately reproduced, while
 $J_1/J_2$ and $i$ are also reliably recovered for both aliased and non-aliased solutions. 
However major error is clearly present for $r_1/r_2$ and
$L_1/L_2$ from aliased solutions. 
Aliasing is the dominant source of systematic error for the data set as a whole. 
The ratio of radii is reasonably well reproduced in non-aliased solutions.
The eccentricity is often not recovered, and has unrealistic standard errors because 
the error is completely dominated by the assumption that $\omega$
 is $0$ or $\pi$. The fitted $e$ is a lower limit.

Fig.~\ref{f10} shows $F_e$ for the simulated detached solutions vs. 
known $F_e$, with
regions of complete and partial eclipse distinguished for both the simulated binaries 
and their solutions. Due to aliasing errors in $i$ and $r_1/r_2$, the solution $F_e$'s 
 are often very poorly reproduced. Most notable are the degraded results for detached binaries
compared to the \emph{SD}'s of Fig. 5. 
However, solutions with $\Delta r_1/r_1<0.05$ and 
$\Delta r_2/r_2<0.05$ (large dots) reliably determine the condition of complete or
partial eclipse.

\subsection{\emph{SD}'s and the aliasing problem}
\label{aliasing}

\noindent The statistics of the aliasing problem, in which solutions with interchanged
radii fit about equally well, are effectively shown by Figs.~\ref{f3}, \ref{f4}, and \ref{f5}
(\emph{SD} solutions of \emph{SD}'s) and Figs.~\ref{f8}, \ref{f9}, and \ref{f10} (detached solutions of
detached binaries). Clearly the problem is considerably less serious for \emph{SD}'s
than for detached systems, and constitutes a further important advantage for the selection
of \emph{SD}'s to be used as ideal distance indicators. 
This is because, while the relative star sizes may appear equally well determined
in the correct and aliased solutions, the luminosity ratio can be very different for systems with unequal
radii and temperatures. Therefore, until high-quality follow up photometry is obtained,
selection of systems with moderate luminosity ratios (to facilitate observation
of double-lined spectra) can be made confidently only for \emph{SD} binaries.

We now ask if that situation can be 
understood in simple terms. Notice that although the members of a well-detached binary 
may differ in surface brightness, \textit{they are twins in figure} (both nearly 
spheres) and, with regard to the solution, \textit{mass ratio is irrelevant}. In particular, 
contributions to ellipsoidal
variation in a well-detached binary are small and comparable for the two stars,
so interchanging the radii has nil effect on
ellipsoidal variation. In \emph{SD}'s, however, the contact component has practically
all of the tidal distortion, so the ratio of radii (which controls the ratio of
luminosities) can have some importance. For example, if we
make the contact star larger, it will necessarily be more luminous and impress more
ellipsoidal variation on the light curve. Figs.~\ref{f3}, \ref{f4}, and \ref{f5}
demonstrate that the difference in ellipsoidal variation is sufficient 
to reduce the severity of aliasing in the \emph{SD} case.

\subsection{the relative quality of \emph{SD} and detached solutions for survey quality data}

\noindent Solutions for $r_1+r_2$ and $J_1/J_2$ are of comparable quality for
\emph{SD} and detached binaries. The solutions for $r_1$ and $r_2$, 
and therefore $r_1/r_2$ and $L_1/L_2$ for detached systems suffer from aliasing,
while these quantities are more reliably recovered for \emph{SD} binaries. In Paper
I, the degeneracy between $i$ and $r_1/r_2$ for partially eclipsing detached 
binaries was discussed in detail. The more accurate recovery of $i$, aside from 
aliasing effects, indicates 
that this degeneracy is not as strong for \emph{SD} binaries. Furthermore,
recovery of $i$ and $r_1/r_2$ leads to the accurate recovery of $F_e$. Thus 
Figs.~\ref{f4}, \ref{f5}, \ref{f9} and \ref{f10} 
demonstrate those two further important advantages for the selection of \emph{SD} systems
that will be suitable for distance indication. For given light-curve quality, 
identification of complete eclipses is more robust for \emph{SD}'s. 
Complete eclipses are important for both detached and \emph{SD} binaries
since they lead to accurate, robust solutions, and hence an accurate
distance. For \emph{SD}'s, this condition can be immediately identified with good reliability, 
without having to obtain further, high quality light-curves. In addition $L_1/L_2$, which governs 
whether the system will have single or double-lined spectra, can be more accurately predicted.

\subsection{can light curves distinguish \emph{SD}'s from detached binaries?}
\label{mimic_solutions}

\begin{figure*}[htbp]
\epsscale{.95}
\plotone{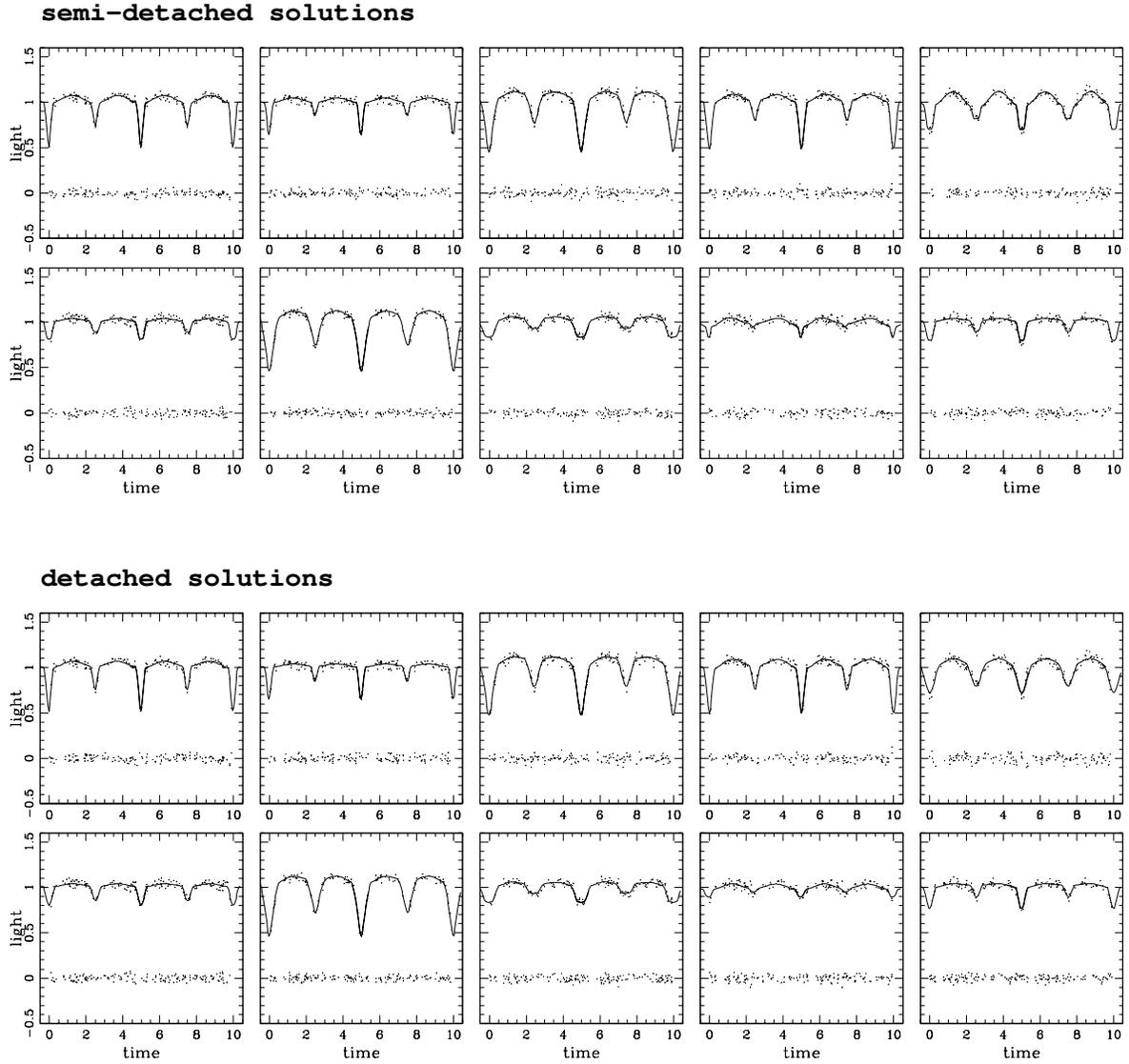}
\caption{\label{f1}Examples of the simulated \emph{SD} light curve 
data with corresponding solutions and residuals. Top: The 
solutions with {\em DC} in mode 5 (for \emph{SD} condition). 
Bottom: The solutions with {\em DC} in mode 2 (for detached 
condition).}
\end{figure*}

\begin{figure*}[htbp]
\epsscale{1.0}
\plotone{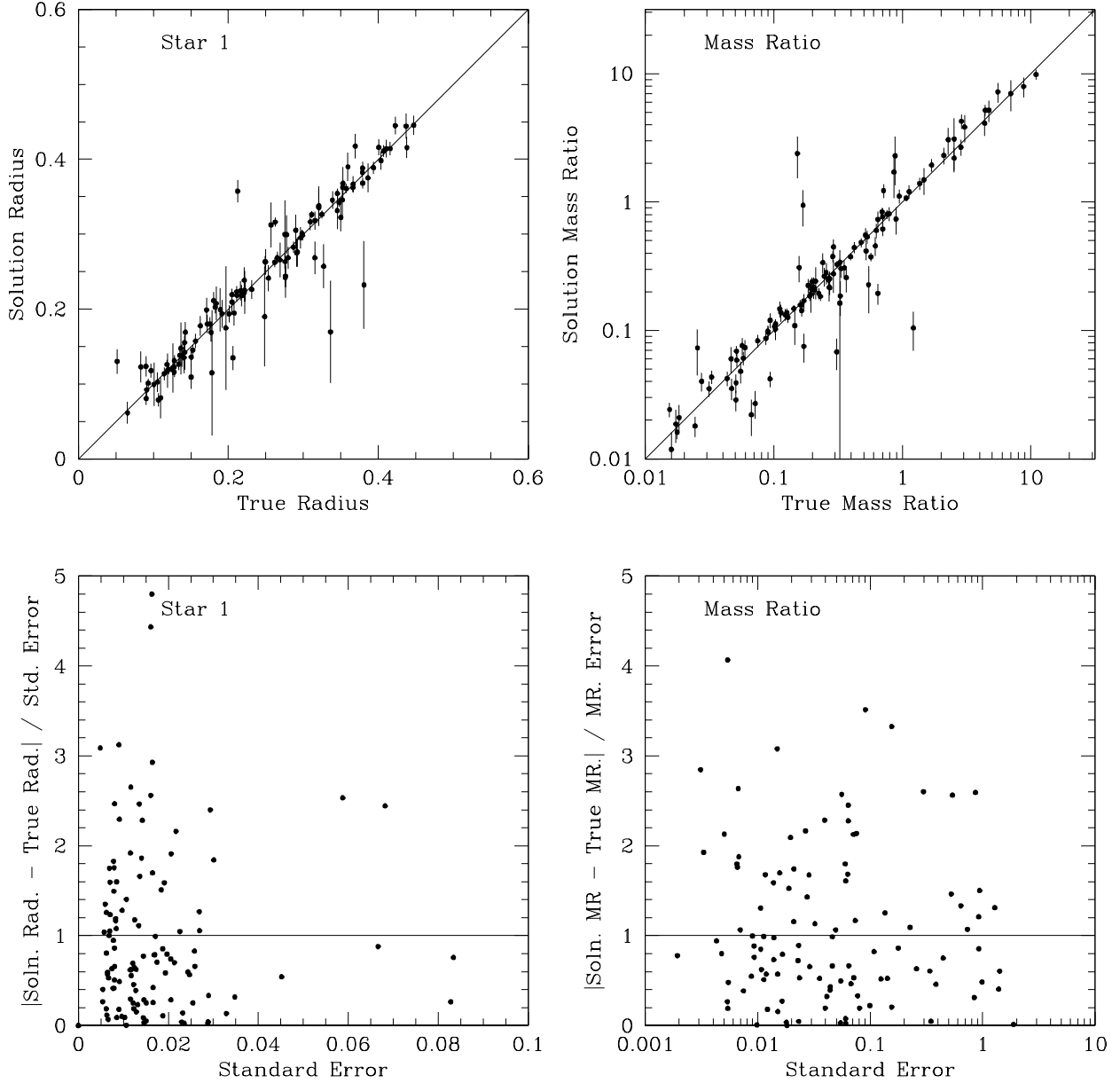}
\caption{\label{f3}Top:  $r_1$ (left)
 and $q$ (right) (with standard errors $\Delta r$ and $\Delta q$)
 from \emph{SD} light curve solutions plotted against $r_{sim}$ and $q_{sim}$ for
 simulated \emph{SD} binaries. The line of equality is
 drawn to guide the eye. Bottom left: $|r-r_{sim}|/\Delta r$ vs. 
 $\Delta r$. Bottom right: $|q-q_{sim}|/\Delta q$
vs. the standard error $\Delta q$. }
\end{figure*}

\begin{figure*}[htbp]
\epsscale{.75}
\plotone{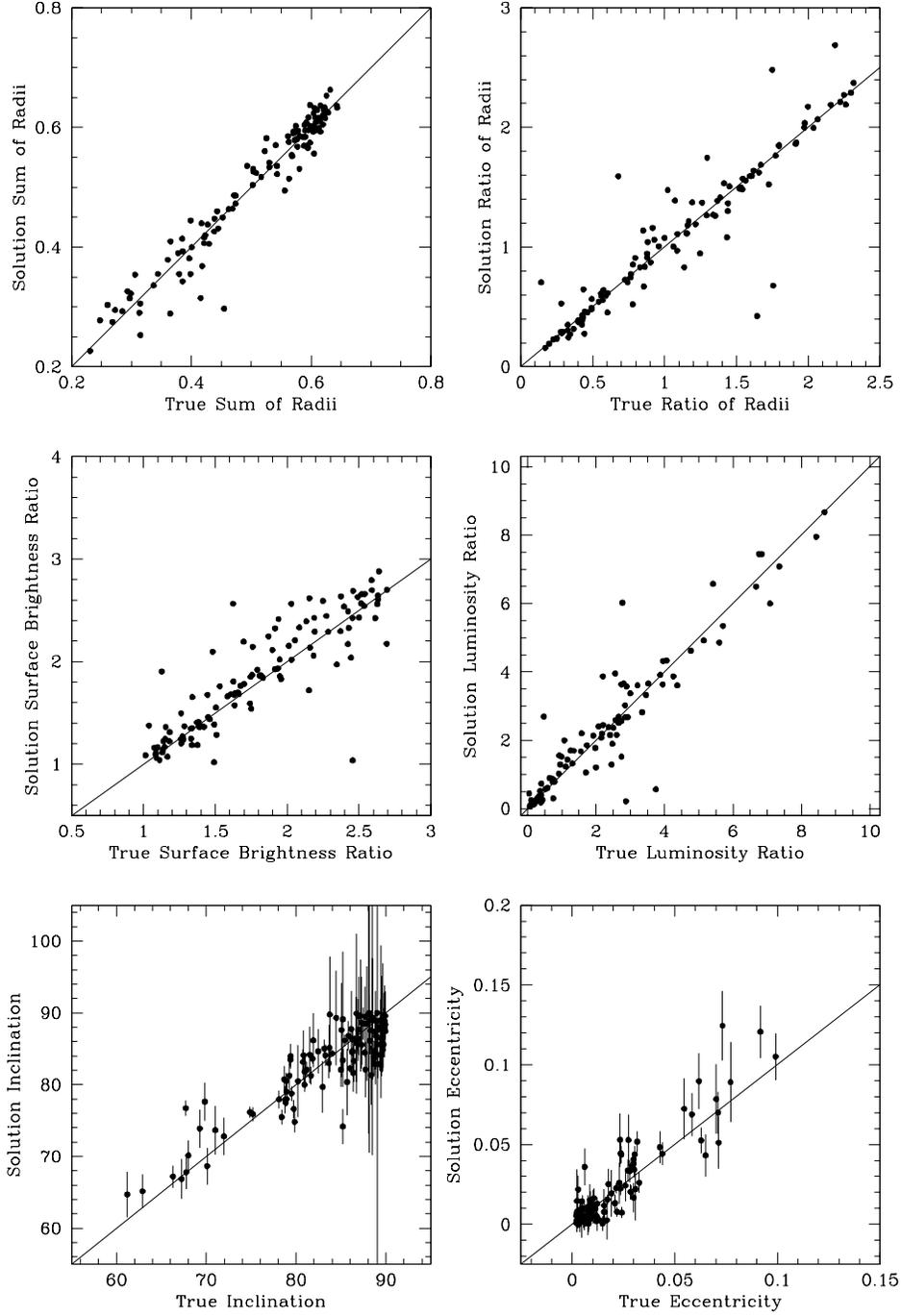}
\caption{\label{f4} \emph{SD} light curve parameters and combined quantities 
 vs. simulated known values for simulated \emph{SD} binaries.
 Plots are shown for $r_1+r_2$,
$r_1/r_2$, $J_1/J_2$, $L_1/L_2$,
$i$, and $e$.}
\end{figure*}

\begin{figure*}[htbp]
\epsscale{.50}
\plotone{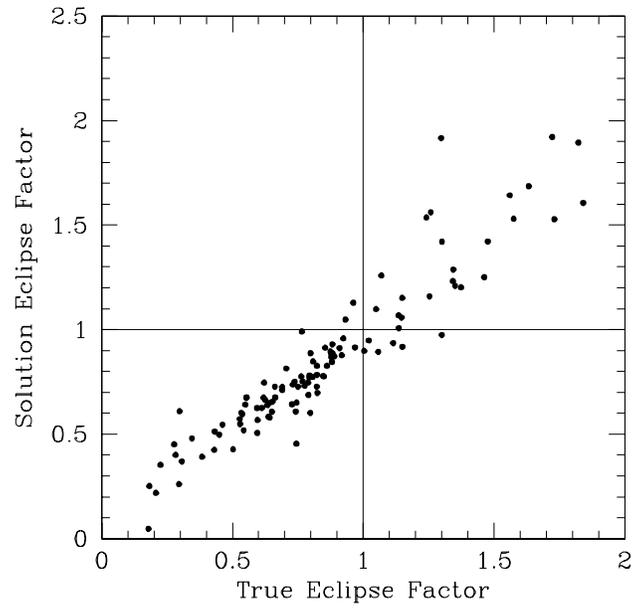}
\caption{\label{f5}$F_e$ from \emph{SD} light
 curve solutions vs. that for the simulated \emph{SD}
 eclipsing binaries, $F_{sim}$. The lines of unity are also drawn to
 separate regions of complete and incomplete eclipse.}
\end{figure*}

\begin{figure*}[htbp]
\epsscale{.95}
\plotone{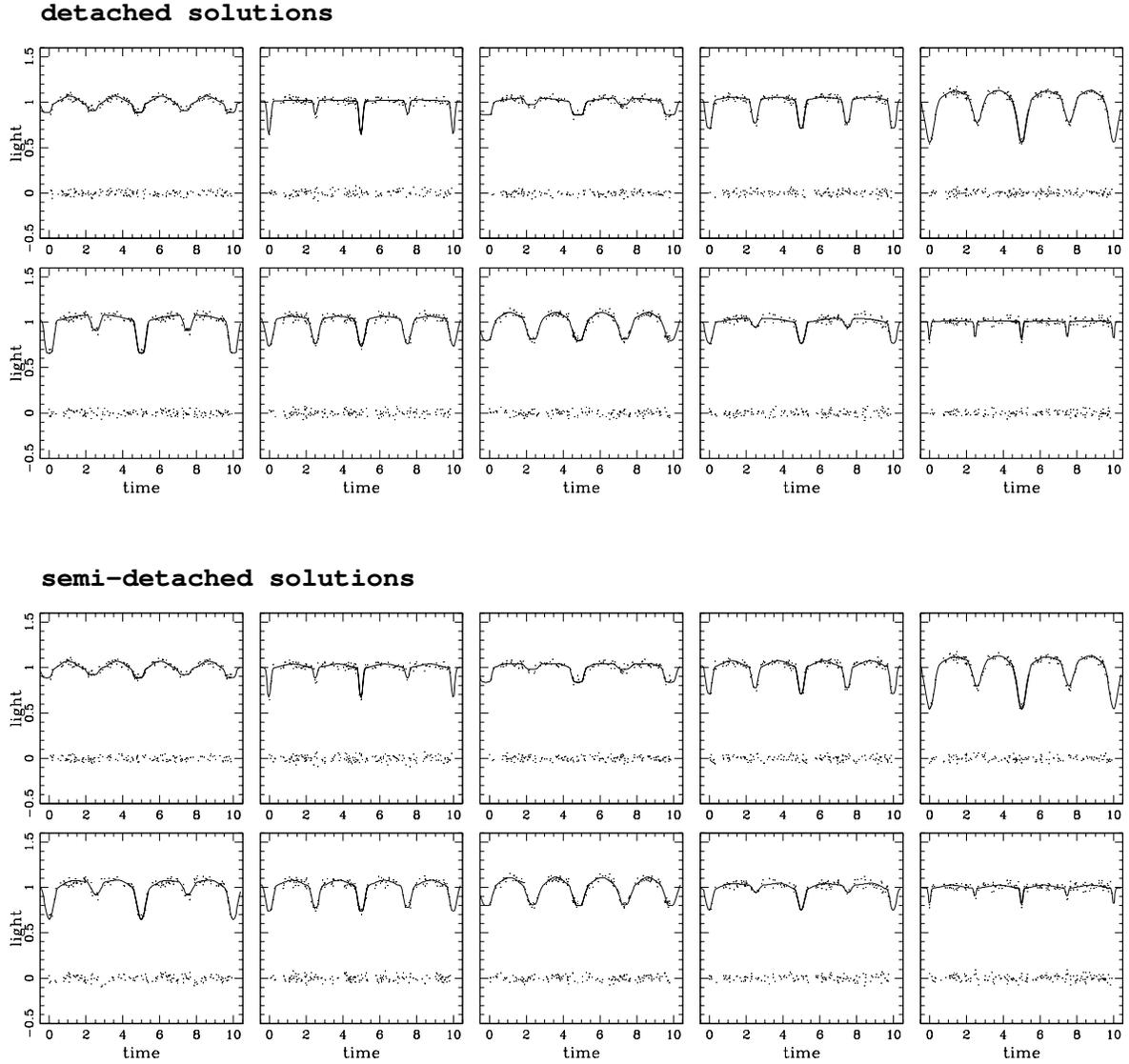}
\caption{\label{f6}Examples of simulated detached light curve 
data with model light curves and residuals. Top: 
Solutions with {\em DC} in mode 2 (for detached condition). 
Bottom: Solutions with {\em DC} in mode 5 (for \emph{SD} 
condition).}
\end{figure*}

\begin{figure*}[htbp]
\epsscale{.85}
\plotone{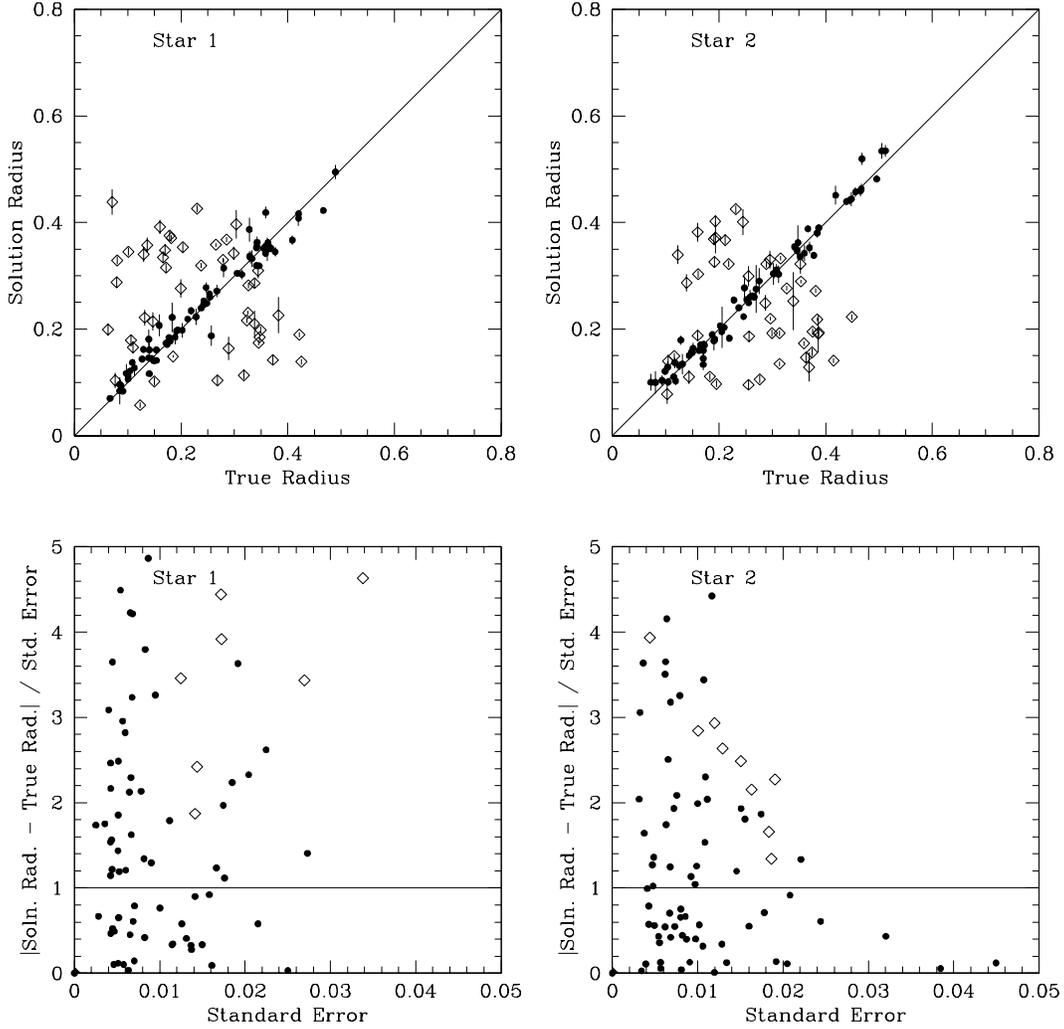}
\caption{\label{f8}Top:  $r_1$ (left)
 and $r_2$ (right) (with standard error, $\Delta r$)
 from light curve solutions vs. $r_{sim}$ for
simulated detached binaries. The line of equality is also
 drawn. Bottom: $|r-r_{sim}|/\Delta r$ vs. 
standard error $\Delta r$. (left: Star 1, right: Star 2). Diamonds show points for which
the ratio of radii is inverted with respect to the correct value, and
both radii differ significantly from the known values (aliased solutions).}
\end{figure*}

\begin{figure*}[htbp]
\epsscale{.75}
\plotone{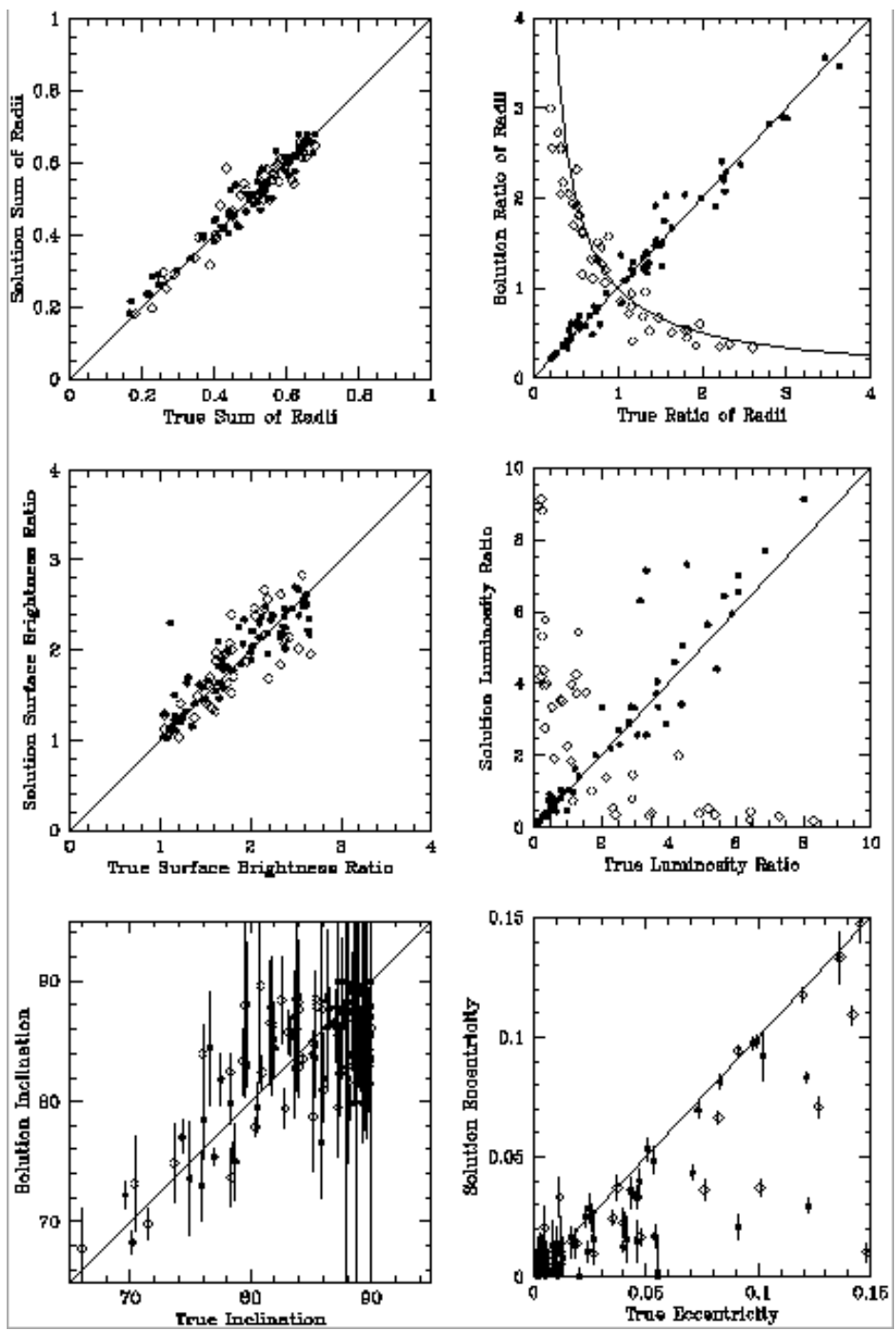}
\caption{\label{f9} Light curve solution parameters with standard
 errors vs. known correct values for simulated detached binaries.
Plots are for $r_1$ vs. $r_2$,
$r_1/r_2$, $J_1/J_2$, $L_1/L_2$,
$i$, and $e$. Diamonds show points where the ratio of radii is inverted with respect
to the simulated system and both radii differ significantly from known values 
(aliased solutions).
The \emph{aliased} line $y=1/x$ is shown to guide the eye.} 
\end{figure*}

\begin{figure*}[htbp]
\epsscale{.5}
\plotone{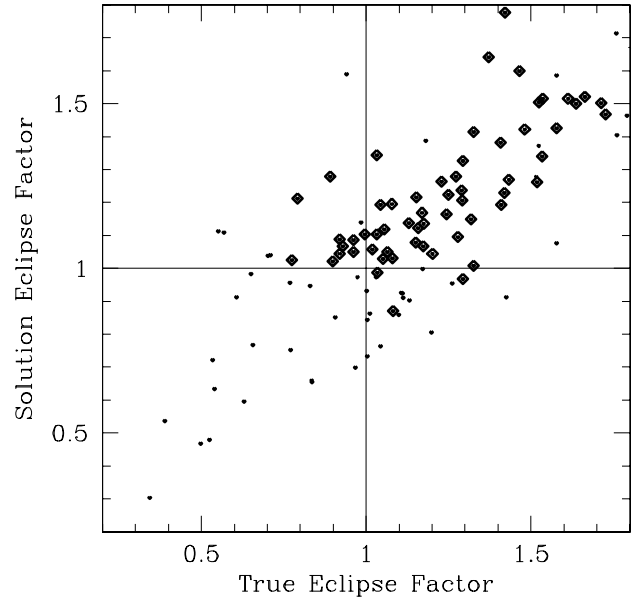}
\caption{\label{f10}$F_e$'s from the light
 curve solution vs. those for simulated detached
 binaries ($F_{sim}$). The lines of unity are drawn to
 separate regions of complete and partial eclipse.
 Cases where $\Delta r_1/r_1<0.05$ and  $\Delta r_2/r_2<0.05$ 
are denoted by larger dots.}
\end{figure*}

\noindent \emph{SD} light curves can resemble those of detached systems and vice-versa, as
 demonstrated in the lower two rows of Figs.~\ref{f1} and \ref{f6}.
In an effort to quantify the problem, the simulated \emph{SD}/detached
systems were fit in reversed modes (detached /\emph{SD})
(Fig.~\ref{f1} / Fig.~\ref{f6}), in addition to the normal way.
A detached solution was obtained for 107 of 120 simulated \emph{SD}'s. 
Mode 2 can mimic the light-curve of an \emph{SD} system, but typically 
$\Omega_2$ is very nearly critical. A small adjustment
can therefore move the solution outside the physically allowable range before 
convergence, resulting in failure to obtain a solution. 
Aliased solutions were found as for the simulated detached binaries. The 
detached solutions can both over and under-estimate $r_2$. However 
the detached solutions tend to overestimate the sum of the polar radii systematically. 
This is expected if the solution essentially finds a correct \textit{mean} radius for
the lobe filling star, as detached stars are less distorted than lobe filling components. 
There appears to be no systematic dependence on solution mode for $r_1/r_2$, $i$ or $J_1/J_2$. 
In the converse situation, \emph{SD} solutions were obtained for 109
of 120 simulated detached systems. In this case, failures tend to result from convergence 
to a solution with residuals larger than the scatter. The \emph{SD} solutions again
can both over and under-estimate $r_2$. There was
no obvious systematic dependence on solution mode for $r_1+r_2$, $r_1/r_2$, or $J_1/J_2$. 
However the \emph{SD} solutions tend to underestimate $i$ systematically for detached binaries. 

Fig.~\ref{f13} illustrates success in determining whether a system 
is \emph{SD} or detached, based on the light-curve residuals. 
Each simulated binary (from both the \emph{SD} and detached catalogs) 
has a point in one (and only one) of the three panels, and simulated \emph{SD} 
and detached binaries are marked by diamonds and dots respectively. 
Simulated binaries with both \emph{SD} and detached solutions are  
in the central panel, where the ratio of the sum of the squares of residuals 
(detached/\emph{SD}, hereafter the \textit{SS} ratio) is plotted against $r_1+r_2$.
 A reasonable fraction (34\%) of systems with both \emph{SD} and detached solutions
have \textit{SS} ratios outside the region of statistical overlap and can be 
reliably classified on that basis.  The lower panel shows 10 simulated systems 
having only a detached solution and 2 misclassified objects. Similarly the top panel 
 shows 13 simulated systems having only an \emph{SD} solution and 3 that have 
been misclassified.

It is important to emphasize that later high quality data from large telescopes will
allow recognition of \emph{SD}'s much more reliably than do OGLE
data, from which we seek only \textit{candidates} for follow-up observations.
A consistency check will exist where double-lined spectra are available, for then one
can compare $q_{ptm}$ from \emph{SD} light curve solutions with $q$ from radial 
velocities ($q_{rv}$). One expects that $q_{ptm}$ will be smaller than $q_{rv}$ for 
detached systems (the detached $q_{ptm}$'s being wrong) and that
the two $q$'s will agree for \emph{SD}'s. There is some help from \emph{SD} solutions
even for binaries with single-lined spectra, because a good $q_{ptm}$ should lead to 
plausible absolute masses for an \emph{SD} system and implausible ones for a detached system.
For examples of high quality data in programs dedicated to individual Magellanic Cloud
eclipsing binaries (binaries observed with a 2.15 m.
telescope), see Ostrov, Lapasset, and Morrell (2000) and Ostrov (2001). One can expect even
better data collections with the larger telescopes and future intensive observing 
justified by the Magellanic Cloud distance problem.
Ribas, \textit{et al.} 2000 similarly
treated the bright LMC eclipsing binary HV 2274 with a telescope of only 0.61 m. aperture.

\section{OGLE\ SMC Binaries - Recognition and Solutions}
\label{OGLEfits}

\noindent The OGLE catalog was recorded automatically and is therefore essentially
free of selection effects, except for those imposed by limiting magnitude
($I\lessapprox 20^{m}$), by period range ($0.^{d}3$ to $250^{d}$), and by
precision (which depends on brightness, \textit{viz. }Udalski, et al.,
1998). Most important is that OGLE stars were not selected according to
observer interest, observational convenience, or horizon location.
Accordingly the catalog is an important new resource not only for distance
estimation but also for binary star statistics, as are the EROS and MACHO catalogs.
However those two purposes
are linked because distance estimation requires reliable recognition of
category membership, which in turn should be checked against expected
category statistics. The first 30 systems from OGLE field 2 are illustrated
in Fig.~\ref{f14} (with \emph{SD} solution fits), so as to give an impression of typical
light curves rather than selected ones. An even better overview can be
attained by inspection of the 1459 light curve panels in Udalski et al.~(1998).  

The most easily recognizable \emph{SD} binaries in our Galaxy are of the Algol
type, such as Algol, TW Draconis, U Sagittae, and S Equulei, to mention a
few. A distinction is made among classical Algols, short-period Algols, and
''pseudo-Algols'' (our term).  Classical Algols are understood to have
experienced an episode of rapid and large scale matter transfer that has now
ended, followed by a long-lived \emph{SD} state of slow transfer. Compared to many
other categories of extrinsic and intrinsic variable stars, they form a
category with very good evolutionary coherence. These are the only
''Algols'' for which one can assume an \emph{SD} state with confidence. The short
period Algols ($P\lessapprox 1$ day) are a mixed bag of uncertain status,
containing some classical Algols, some slightly detached main sequence
binaries, and even some erstwhile \emph{OC} systems that have temporarily
broken contact. The third category, pseudo-Algols, is primarily composed
of main sequence detached binaries with
very unequal (primary vs. secondary) temperatures, although a minority are
bizarre products of moderately advanced evolution. That is, pseudo-Algols
are simply mis-typed, although some catalogs of Algols have them in large
numbers, as their light curves superficially resemble those of (classical)
Algols. Evolved pseudo-Algols are likely to be \emph{SD}, but with more
complicated histories than classical Algols.  

The signature light curves, by
which all of these categories are assigned the name \textit{Algol type,}
have deep primary and shallow (in extreme cases, nearly undetectable) secondary
eclipses. There usually is some variation between eclipses due to the ``reflection'' effect
and tidal deformation, but typically with  $\lessapprox
0^{m}.1$ amplitude. 
Among the three ``Algol'' categories, classical Algols conform
best to the above light curve description and can thereby be rather reliably
recognized, provided that the inclination is high (say $>80^{\circ }$) and
the period is longer than a day or so. 

Curiously, remarkably few OGLE light curves resemble those of the common
classical Algols, which should be abundant according to experience with
Milky Way binaries. Typically, the OGLE secondary eclipses are comparatively
deep or there is too much variation between eclipses, or both conditions
occur. Reasons to consider are:

\noindent $\bullet$ Could Milky Way binary statistics be biased in favor of finding
''normal'' Algols, perhaps observationally or according to the interests of
observers? Might there even be a bias according to what becomes published?
Unfortunately, existing statistical compilations do not reliably distinguish
classical Algols from Algol look-alikes. Resolution of these issues would be
an interesting project, but is beyond the scope of this paper. 

\noindent $\bullet$ Could OGLE SMC statistics be biased against finding ''normal'' Algols?
Certainly there is no bias against period, as
Algol periods lie well within OGLE limits. Algols with primaries of spectral
types middle A and later might be too faint for the OGLE limiting
magnitudes, but B-types should be readily observable. Detection of the deep
eclipses of Algols is assured, given OGLE's precision. Therefore, although this
possibility might explain a substantial deficiency of Algols, it would not seem to explain
their near absence in OGLE. By eye we see only about 20 of the 1459 light curves that might
resemble those of common Algols, and only about 5 appear entirely normal.

\noindent $\bullet$ Could we be misled by the printed OGLE light curves being for the \textit{I}
band, while most light curves in journals are for \textit{V} or \textit{B}?
Indeed, \textit{I} band curves will have deeper secondary and shallower
primary eclipses, as well as increased ellipsoidal variation, compared to
shorter wavelength bands. Although these effects are in the right sense, one
can make approximate allowances, and they seem insufficient to account for
OGLE's apparent shortfall in normal Algol light curves.

\noindent $\bullet$ Could Algols be far less abundant in the SMC than in the Milky Way? At
first sight this seems unlikely, but it must be considered, given that
several other kinds of objects differ statistically between the SMC and Milky Way.
For example, Algol formation could be sensitive to chemical composition,
given that a proper physical and quantitative theory of Algol formation does not exist.

\noindent $\bullet$ A minor point is that what appear to be alternating eclipses of 
equal depth may in some cases be successive primary eclipses, with undetectable 
secondaries in between and the period being half of that assumed. However that 
possibility could add only a few Algols, at most.

Note that we do not require that our \emph{SD}'s be
Algols, and certainly not necessarily Algols of the most ordinary kind. It
is just that one expects ordinary Algols to be the most abundant \emph{SD}'s, yet
very few OGLE light curves resemble those of the familiar classical Algols. 

Strong tidal dissipation associated with lobe filling
causes the vast majority of \emph{SD}'s to have circular orbits, so eccentricity is
a very useful practical discriminant by which to filter out non-\emph{SD}'s (lobe
filling for an eccentric binary basically means that the lobe is filled at periastron).
We shall rely on the \textit{SS} ratio test and absence of eccentricity to
identify \emph{SD} systems and defer to later
the issue of how many OGLE \emph{SD}'s can be considered Algols.

\subsection{\emph{SD} and detached solutions for SMC binaries}

\noindent Now having some experience with simulated \emph{SD} 
binaries and with the ''normal Algol'' 
issue put aside, we applied our algorithm to the OGLE catalog and
found \emph{SD} and detached solutions for 92 percent and 88 percent of eclipsing
binaries, respectively. We found solutions of both kinds for 83 percent.

Figs.~\ref{f15}-\ref{f17} give an overall impression of the \emph{SD} solution statistics for all
systems in all the SMC fields and demonstrate the range of solutions and
range of solution quality in plots of $r_{1}$ vs. $r_{2}$ (Fig.~\ref{f15}), $q$ vs. 
$r_{1}$ (Fig.~\ref{f16}), and $L_{1}/L_{2}$ vs. $J_{1}/J_{2}$ (Fig.~\ref{f17}). A dearth
of binaries with $r_{1}\approx r_{2}$ appears as a gap that is not found in corresponding
plots for simulated binary catalogs, and is therefore not an artifact
of the analysis. The gap also appears in
the $q$ statistics, but again not in plots for simulated binary catalogs. Classical 
Algols would be found in the 
upper right hand corner of the plots in Fig.~\ref{f17}. The range of solutions for $q$, 
$L_{1}/L_{2}$, and $J_{1}/J_{2}$ is comparable to that of the simulated \emph{SD}
catalog. 

The range and quality of detached solutions for $r_{1}$ and $r_{2}$ are shown in Fig.~\ref{f19} 
(see also Paper I). The dearth of $r_{1}\approx r_{2}$ \emph{SD}'s in 
Fig.~\ref{f15} is not seen in the
detached results. The range of solutions for $r_{1}$ and $r_{2}$ is comparable to
that of the simulated detached catalog.

\subsection{Selection of candidate \emph{SD} OGLE binaries}

\begin{figure*}[htbp]
\epsscale{.75}
\plotone{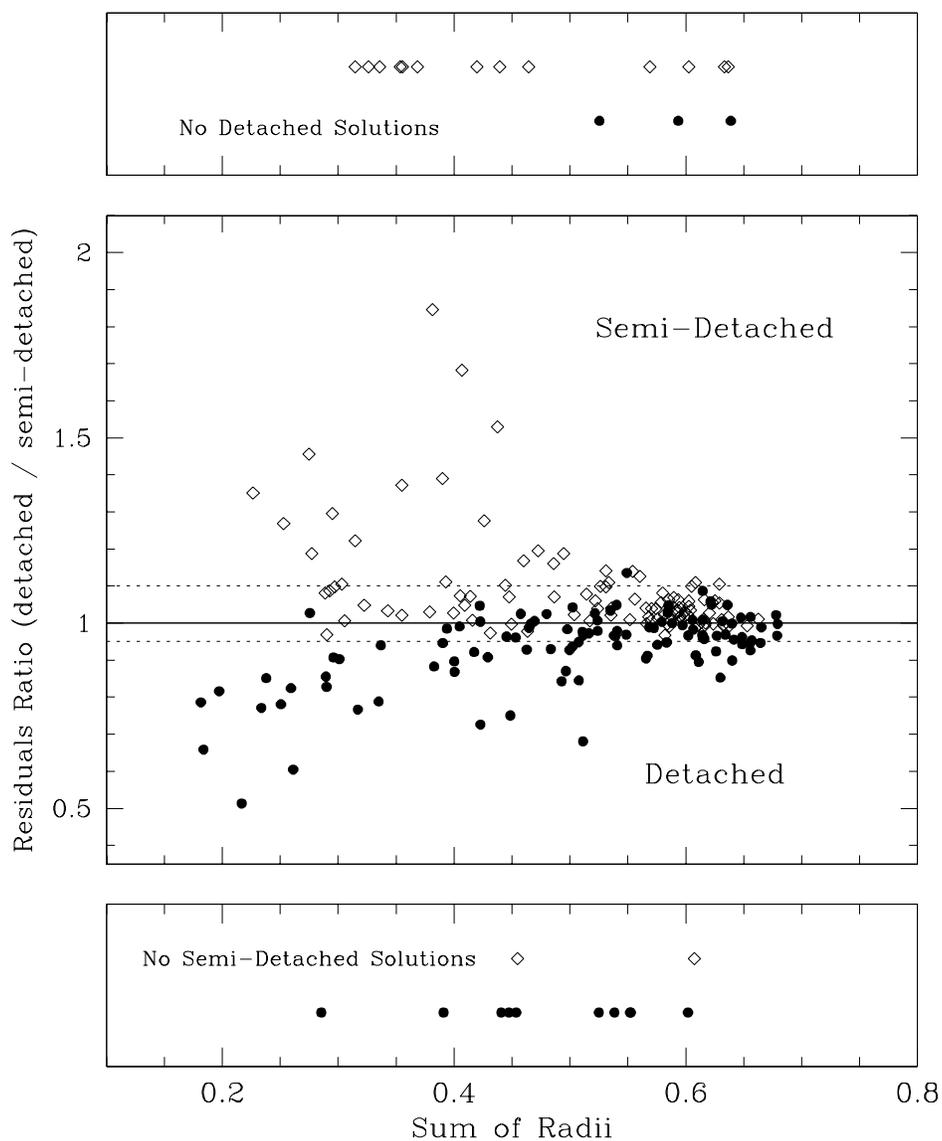}
\caption{\label{f13} Central panel: \textit{SS} ratio for detached
to \emph{SD} catalog solutions vs. $r_1+r_2$. Diamonds are for the \emph{SD}
catalog and dots for the detached catalog. Horizontal dashed lines at 0.95 and
1.10 essentially bound the region of uncertain classification. Lower panel: Simulated
systems having only a detached solution. Diamonds show misclassified objects.
Upper panel: Simulated systems having only an \emph{SD} solution, with dots for 
misclassified objects.}
\end{figure*}

\begin{figure*}[htbp]
\epsscale{.95}
\plotone{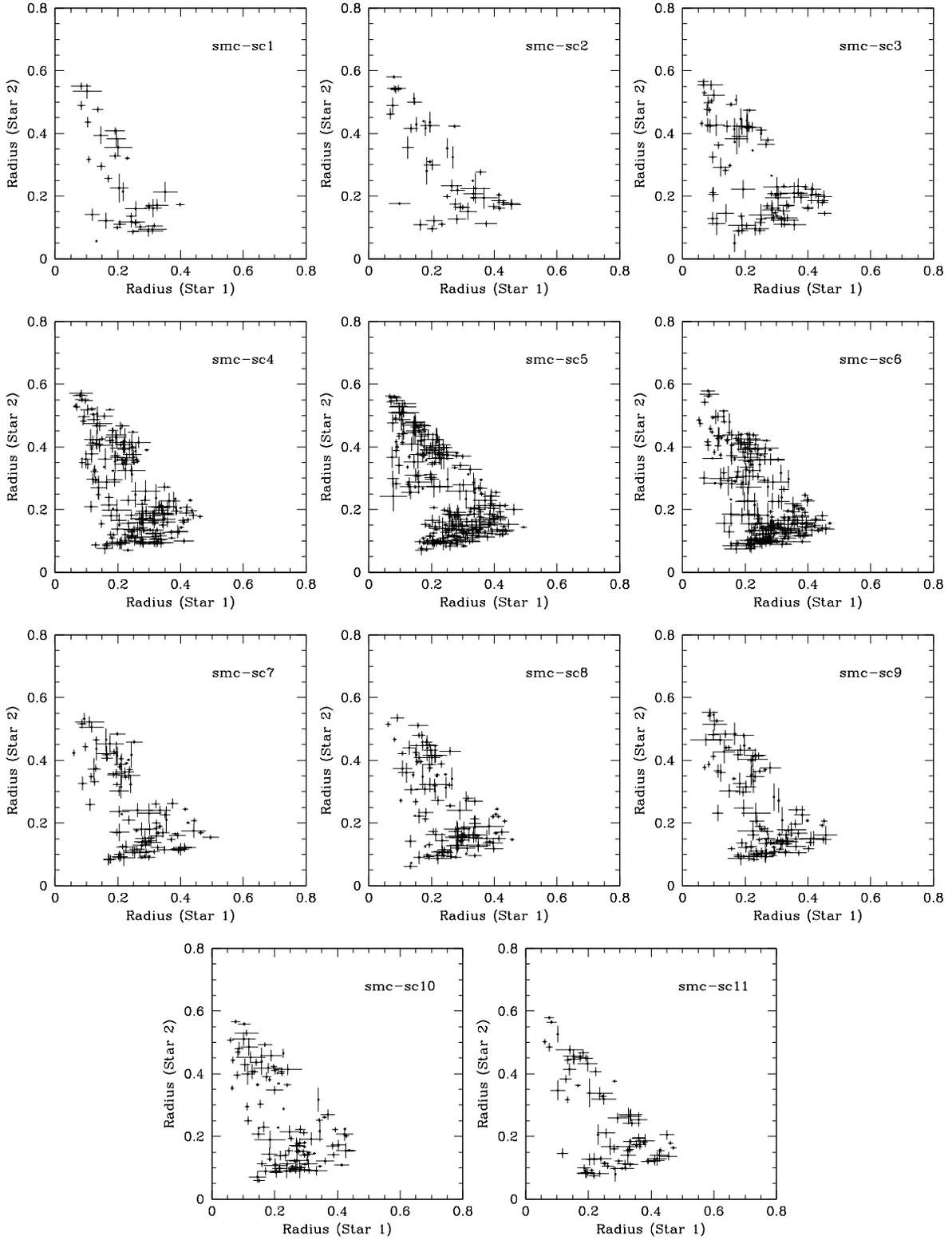}
\caption{\label{f15} \emph{SD} OGLE solutions for $r_1$ vs. $r_2$. Only error bars smaller than 0.05 are shown.}
\end{figure*}

\begin{figure*}[htbp]
\epsscale{.95}
\plotone{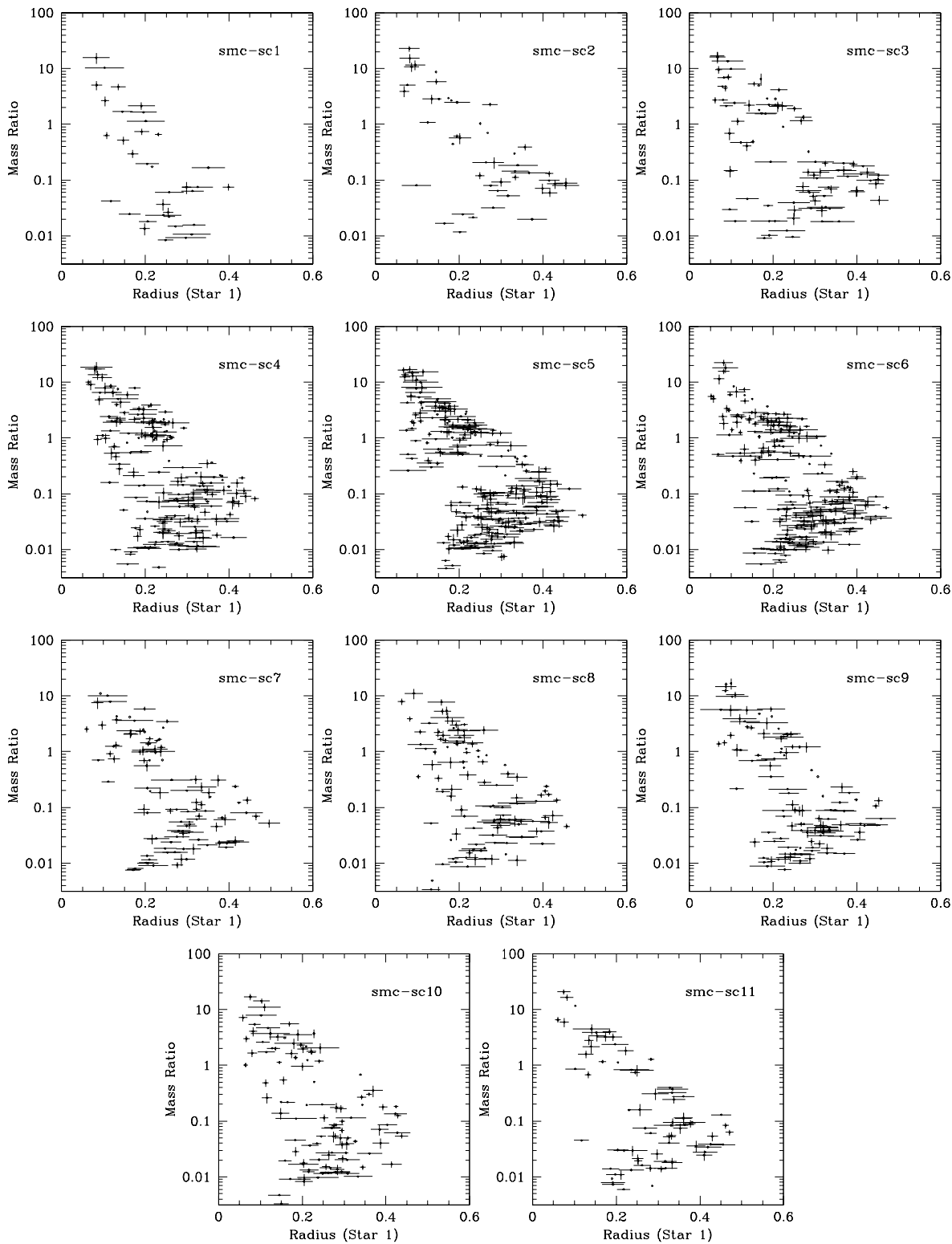}
\caption{\label{f16} $q$ vs. $r_1$ for \emph{SD} OGLE solutions.
Only fractional error bars in $q$ smaller than 0.25, and
in $r_1$ smaller than 0.05, are shown.}
\end{figure*}

\begin{figure*}[htbp]
\epsscale{.95}
\plotone{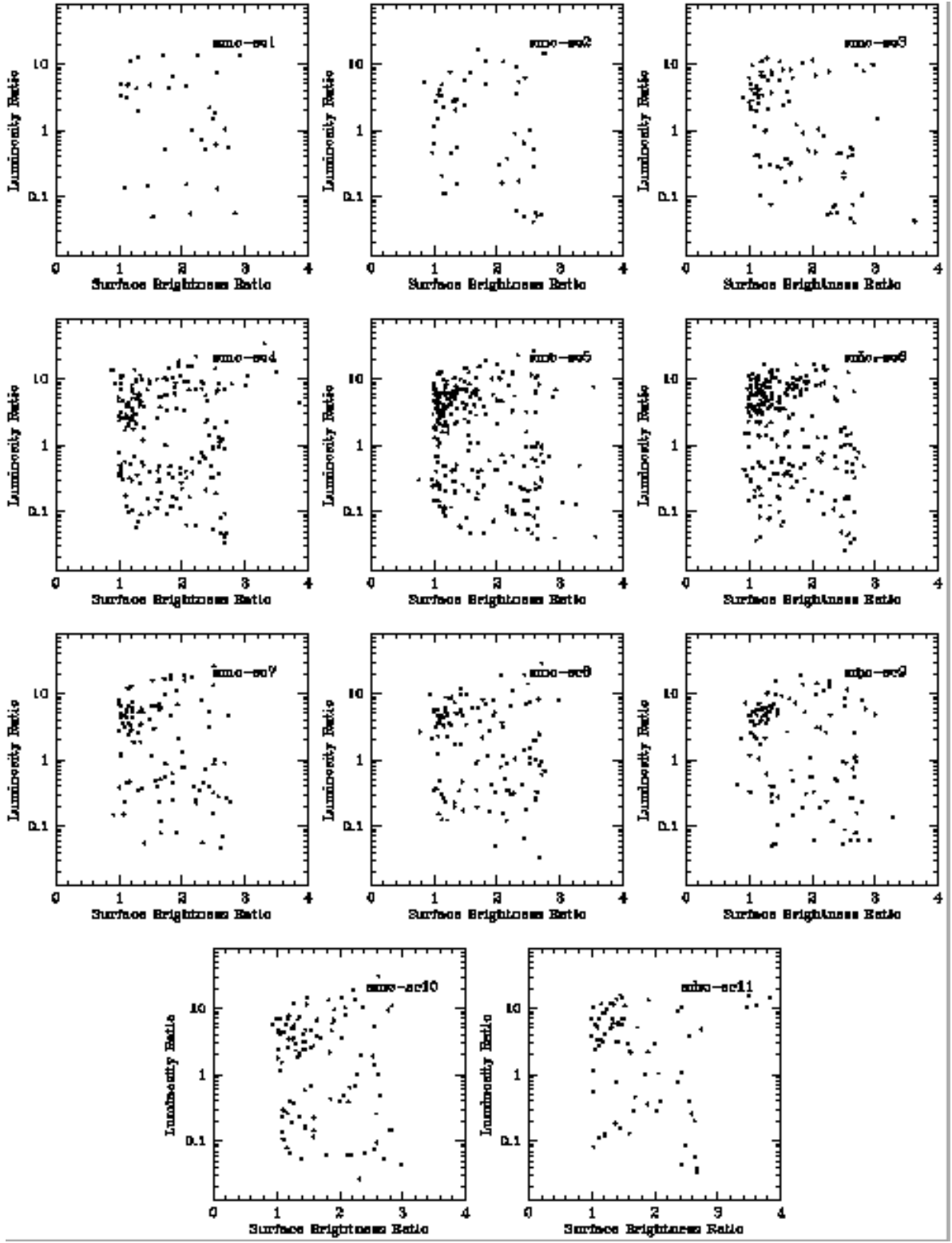}
\caption{\label{f17}\emph{SD} solutions for $L_1/L_2$ vs. $J_1/J_2$.}
\end{figure*}

\begin{figure*}[htbp]
\epsscale{.95}
\plotone{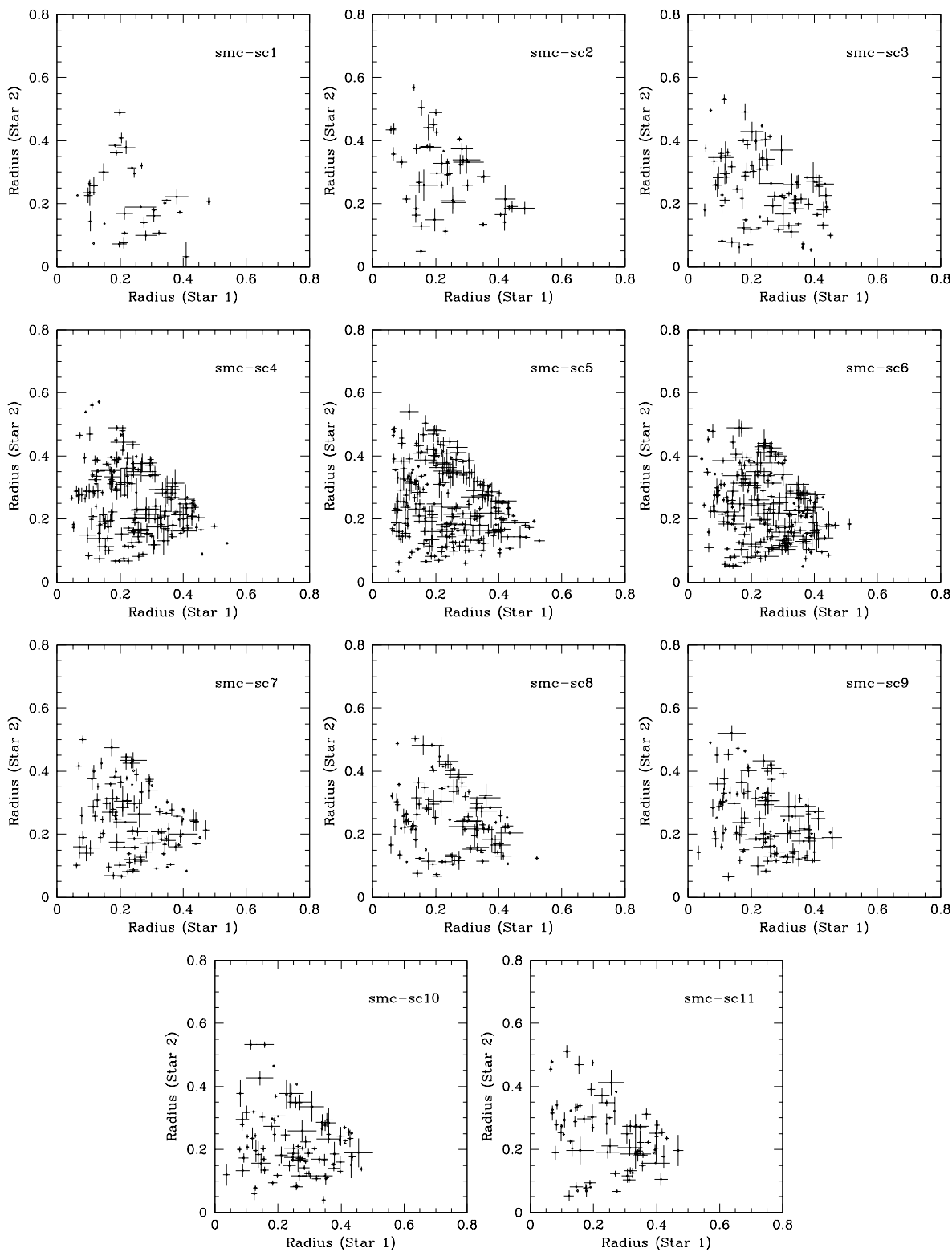}
\caption{\label{f19} Detached solutions for $r_1$ vs. $r_2$. Only error bars smaller than 0.05 are shown.}
\end{figure*}

\begin{figure*}[htbp]
\epsscale{.75}
\plotone{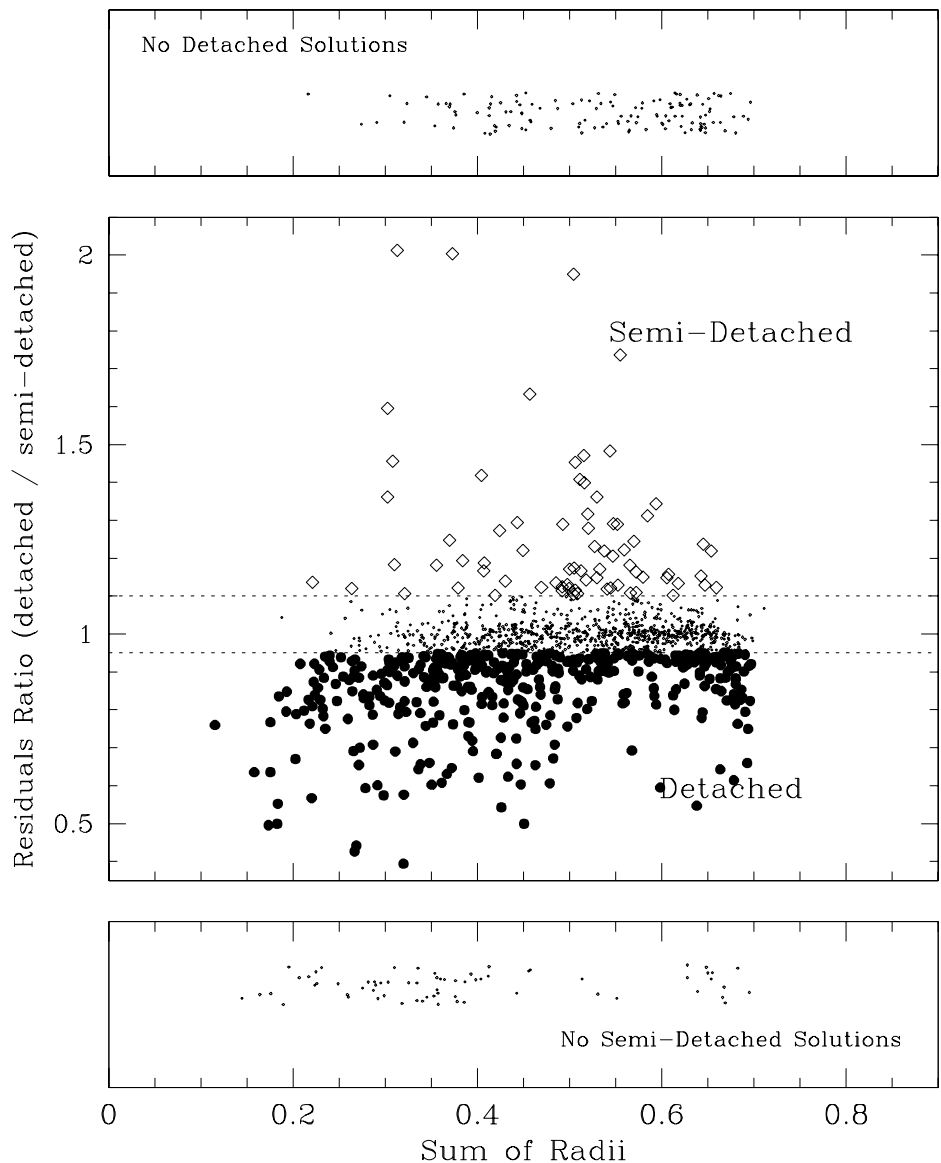}
\caption{\label{f20} Central panel: \textit{SS} ratio for
detached to \emph{SD} solutions of OGLE stars vs. $r_1+r_2$. Diamonds show \emph{SD}
candidates and large dots detached candidates. Horizontal dashed lines at 0.95 and 1.10 delimit
the uncertain region. Lower panel: Systems with only a detached solution.
Upper panel: Systems with only a \emph{SD} solution. The random scatter in the upper and 
lower panels has been introduced for clarity.}
\end{figure*}

\noindent Ratios of \textit{SS} between detached and \emph{SD} OGLE solutions are plotted 
against $r_{1}+r_{2}$ in Fig.~\ref{f20}, corresponding to Fig.~\ref{f13} for the 
simulated binaries. We found
from Fig.~\ref{f13} that systems with a solution in only one mode do not have their
condition reliably predicted by that mode. That is, although the successful
solution mode correctly predicts a system's morphological type in most such
examples, there are too many exceptions for confident assessment. Both main
sequence and RS CVn type binaries may be significant contaminants in
selection of \emph{SD} binaries from the upper panels of Fig.~\ref{f20}. RS~CVn's
are abundant detached binaries that have evolved beyond the main sequence.
The more massive star is larger and cooler and periods range from about 2
days to 2 weeks.

Of the binaries that have solutions in both \emph{SD} and detached modes, 6 percent have 
\textit{SS} ratios greater than 1.1, and 30 percent have \textit{SS}$<0.95$. Our selected 
\emph{SD}'s have solutions in both \emph{SD} and detached mode and \textit{SS} ratios
$>1.1$. Fig.~\ref{f20} shows that our fitting algorithm is more likely to fail in
mode 2 when $r_{1}+r_{2}$ is large, and in mode 5 when it is small. However
solutions are found over a wide range of $r_{1}+r_{2}$. We found no
trend of \textit{SS} ratio with system brightness. In particular, bright binaries are
not preferentially selected as \emph{SD}, based on the \textit{SS} ratio criterion.

\begin{figure*}[htbp]
\epsscale{.95}
\plotone{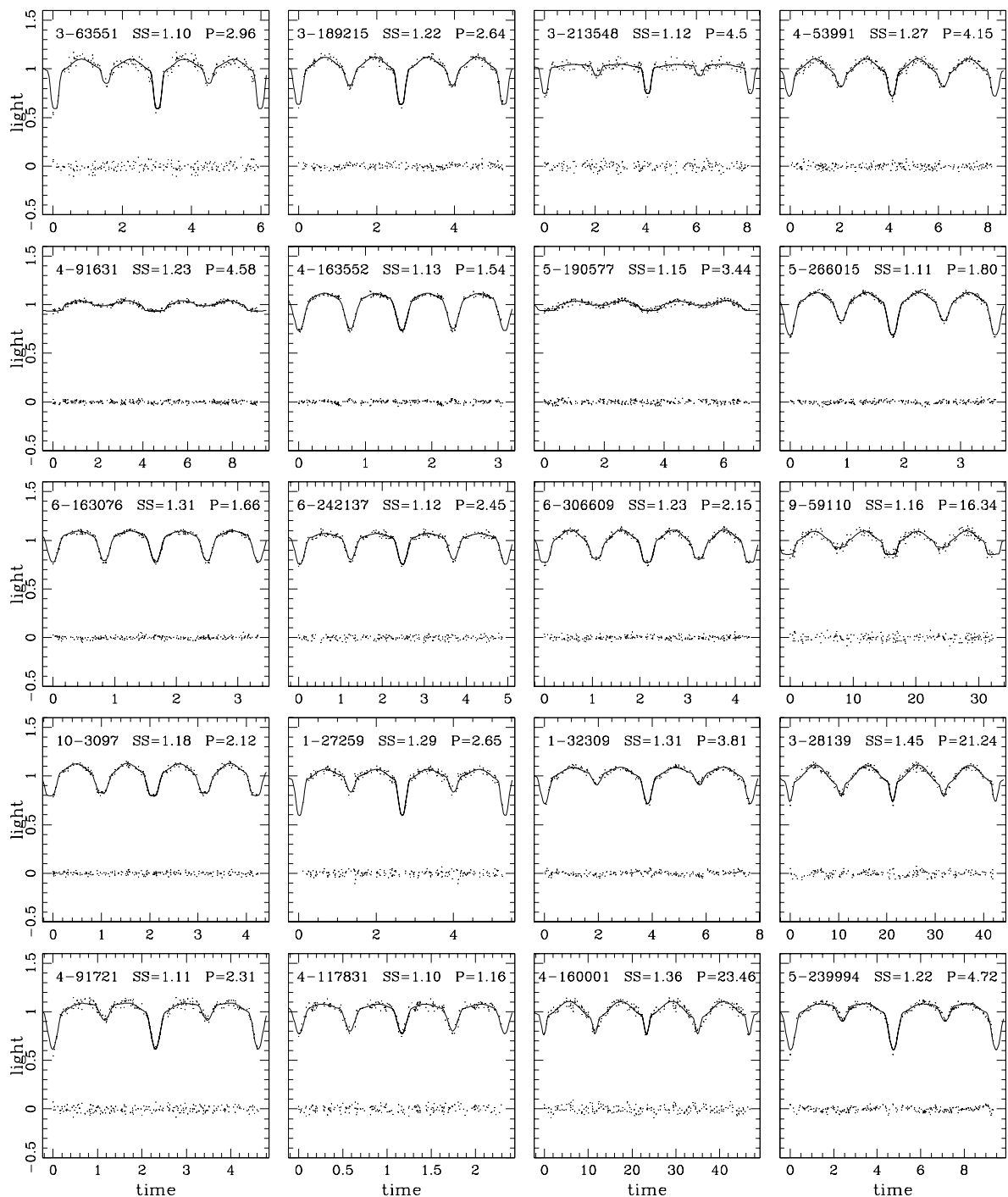}
\caption{\label{f22} Light-curve fits for likely \emph{SD} 
OGLE eclipsing binaries. Each panel is labeled by the OGLE field and object identification 
number, as well as the \textit{SS} ratio and period. The solution parameters are in Tab.~\ref{t4}.}
\end{figure*}

\begin{figure*}[htbp]
\epsscale{.95}
\plotone{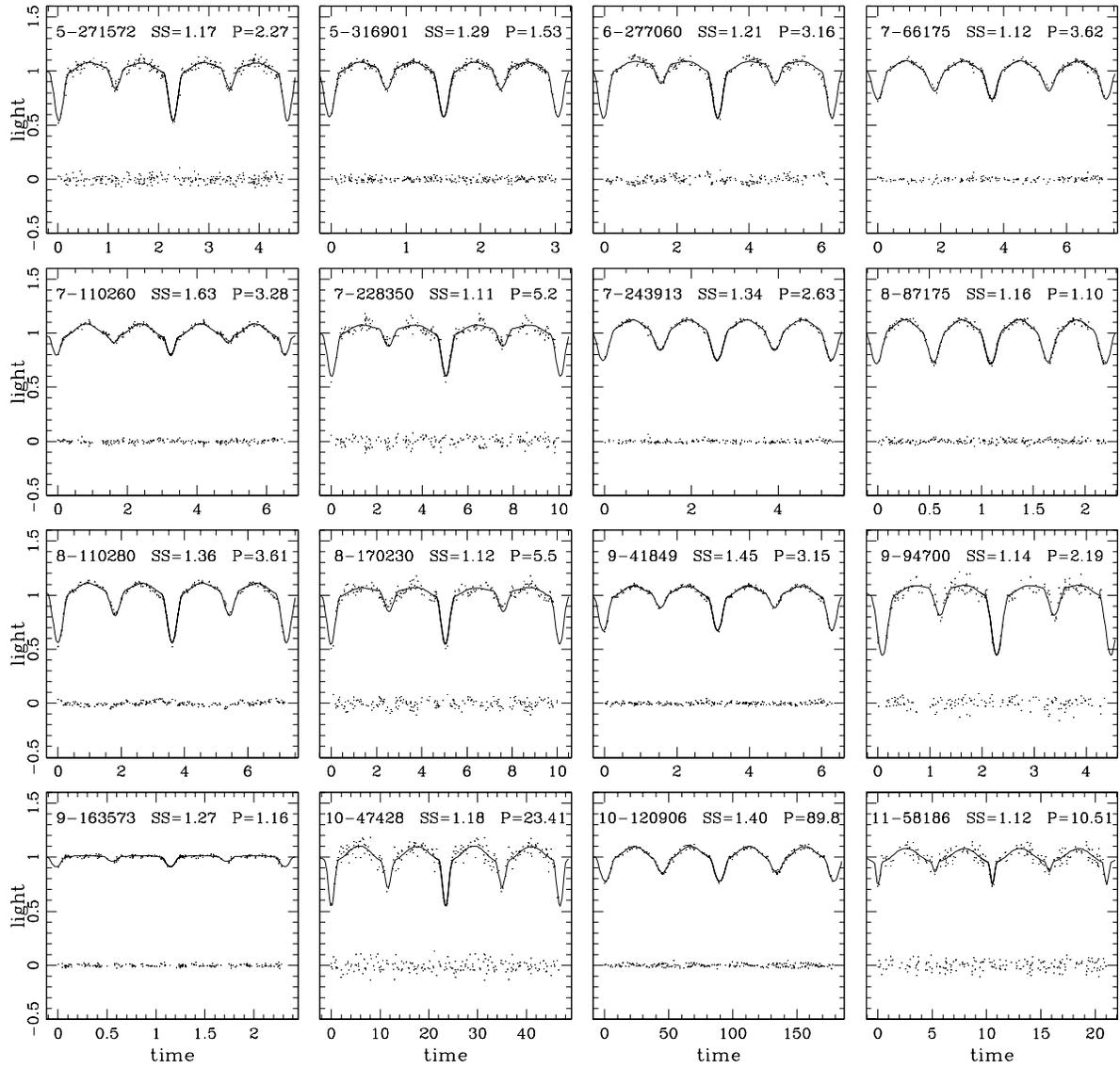}
\caption{\label{f23} Continuation of Fig.~\ref{f22}.}
\end{figure*}

\begin{figure*}[htbp]
\epsscale{1.}
\plotone{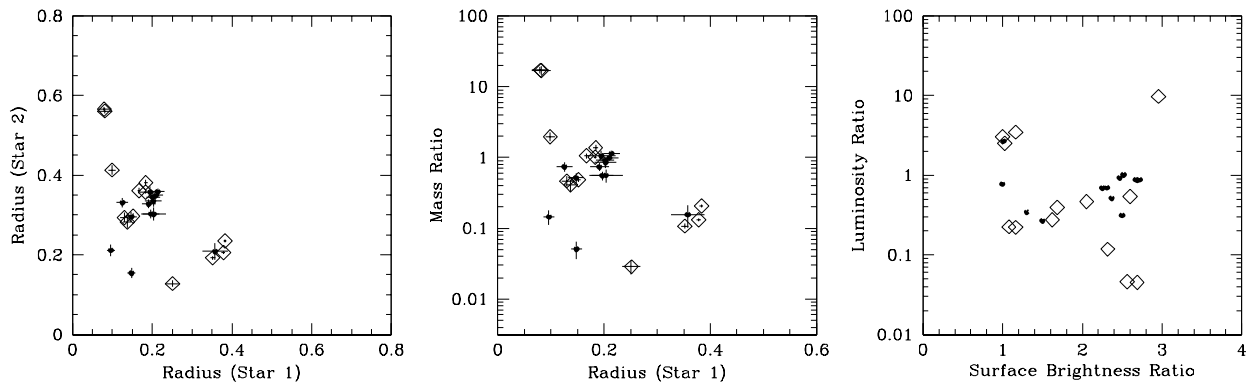}
\caption{\label{f24}  Solutions for the candidate \emph{SD} catalog. 
Left: $r_2$ vs. $r_1$. Center: $q$ vs. $r_1$.
Right: $L_1/L_2$ vs. $J_1/J_2$. Only standard errors in radius smaller than
0.05 and in $q$ (fractional errors) smaller than 0.25 are plotted. The dots and diamonds designate 
partially and completely eclipsing binaries respectively.}
\end{figure*}

Figs.~\ref{f22} and \ref{f23} show \emph{SD} light curve fits for OGLE binaries selected from the
requirement that $e<3\Delta e$ and from the \textit{SS} ratio criterion. Of the objects
having \textit{SS}$>$1.1, 50 percent had $e$ inconsistent with zero. Most of the systems 
rejected on this basis had solutions for mass ratio larger than or around 1, consistent 
with their being detached objects, but having the more massive component evolved, or 
even being of RS~CVn type.
The solutions are summarized in Tab.~\ref{t4}. The first $13$ systems are completely
eclipsing according to the $F_{e}>1$ criterion and are among the best
candidates for distance determination. Fig.~\ref{f24} shows $r_{1}$ vs.
$r_{2}$, $q$ vs. $r_{1}$, and $L_{1}/L_{2}$ vs. $J_{1}/J_{2}$ for systems
judged to have complete (diamonds) and partial (dots) eclipses. The figures and
Tab.~\ref{t4} demonstrate that the instances of complete eclipse result from $r_{2}$
being larger than $r_{1}$, from larger $q$'s, and from larger $i$'s. Most of
the systems have $q$'s of a few $10$ths to $1$, so the primaries will have
significant \textit{RV} amplitudes. 

Both completely and partially eclipsing \emph{SD}'s were selected over a surprisingly 
large range of mass ratio. Systems 4-91631 and 5-190577
have $q>10$, and 9-59110 has $q=1.96$. These have the smallest $r_{1}$'s as well
as the smallest $L_1/L_2$ and could possibly be \emph{SD} binaries in the early 
stages of mass transfer. However the light curves are not erratic, as would be expected 
for rapid mass transfer, so they are more likely to be evolved but detached systems. 
If so these stars are very unusual astrophysically and worth observing as objects
of special interest, aside from distance determination. 

The majority of systems have luminosity ratios of several 10ths to 1, indicating that 
their spectra will be double-lined.
However two objects (3-213548 and 9-163573), have large luminosity ratios, q's 
of $\approx 0.03$ and surface brightness ratios of 3.0 and 1.8. These (particularly
3-213548) are the objects most likely to be classical Algols (as is apparent
from their light-curves, although 9-163573 has very shallow eclipses).
 The near absence of normal Algol light curves in the 
OGLE catalog was discussed at length at the beginning of this section. Their near
absence in our candidate list is therefore not due to the selection criterion. Of the 5
systems with light-curves (selected by eye) that most resemble those of classical
Algols, 4-131230, 8-20567 and 5-283889 had \textit{SS} ratios near 1, and 
5-289333 and 6-200243 had $SS>1.1$, but statistically non-zero $e$'s. 

\begin{figure*}[htbp]
\epsscale{.95}
\plotone{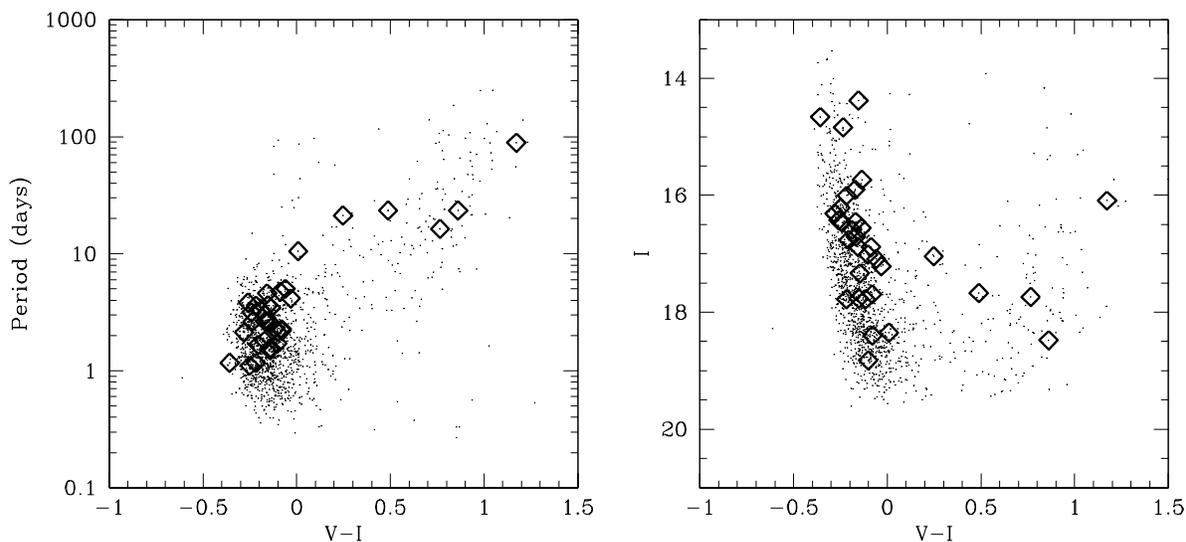}
\caption{\label{f25} Color - period and color - magnitude diagrams for the full eclipsing binary catalog (small dots, data from Udalski et al. 1998). The \emph{SD} objects are superimposed as diamonds. The V-I colors have been de-reddened assuming E(V-I)=0.14.}
\end{figure*}

Fig.~\ref{f25} shows color-period and color-magnitude diagrams for all 36
candidate \emph{SD} OGLE binaries, superimposed on the whole OGLE eclipsing binary
catalog. The candidate \emph{SD} binaries have color, magnitude, and period
distributions similar to those of the whole catalog. Most have colors of
hot main sequence stars. A few have redder colors,
suggesting that the evolved component still contributes a significant
fraction of total light. The two objects with very large $q$'s do not have
overly red colors, but the completely eclipsing system with $q=1.96$ has 
\textit{V-I}$=0.91$. Most of the candidate \emph{SD} binaries have periods of a few
days, consistent with typical Algol periods, although most of the light
curves are not typical of Algols. Those binaries with longer periods also
have redder colors and are the least likely \emph{SD} candidates.

\section{Conclusions}

Although traditionally used, well detached eclipsing binaries 
are not necessarily the ideal or only choice for eclipsing binary
distance determination, since absolute brightnesses on absolute stellar surface
elements can be computed and integrated over surfaces for almost all classes of eclipsing binaries. 
In this paper we have presented arguments in support of semi-detached (\emph{SD}) 
eclipsing binaries as standard candles, and taken this as motivation to find \emph{SD} 
solutions to the OGLE SMC eclipsing binary catalog, and to select 
\emph{SD} systems for future study.

Several advantages of \emph{SD} binaries, in particular the exploitation of lobe-filling 
configurations lead to accurate light-curve solutions and may therefore lead to accurate 
distances. However before investing the considerable effort 
to make the observations required for a distance determination, it is helpful
to have confidence that the binary is appropriate for the purpose. We have found several 
advantages for \emph{SD} systems that relate to the selection of candidates from the 
large catalogs of light-curves now becoming available. We find that aliased solutions 
(where the radii are interchanged) are significantly less of a problem
for \emph{SD} than detached systems. Furthermore, the inclination, and therefore the 
ratios of radii and luminosity, as well as whether the system undergoes complete
eclipse are much better determined. Candidate \emph{SD} distance determination systems
can therefore be selected for the desirable properties of having 
double-lined spectra and complete eclipses more reliably than can
detached systems.

We have computed both \emph{SD} and detached solutions to the 1459 
eclipsing binary stars identified in the SMC by the OGLE collaboration
(Udalski et al.~1998). This work follows our earlier paper on detached systems.
By fitting simulated catalogs we estimate a success rate of 98 percent for finding acceptable
converged solutions to \emph{SD} configurations. Acceptance of an \emph{SD} solution
does not establish that a system surely is \emph{SD}, as detached systems can mimic \emph{SD} 
light-curves and vice-versa. However, we show that the system condition (detached or \emph{SD}) 
can often (for about 1/3 of our simulated binaries) be determined from the ratio of residuals
(\textit{SS} ratio). 
Of the OGLE systems with both kinds of solution, 36 percent have \textit{SS} ratios 
significantly different from unity, allowing morphological categorization.
In particular, also requiring eccentricity consistent with zero, we find 36 systems that
can be identified as being of the \emph{SD} morphological type with reasonable reliability 
(although 7 of these are doubtful due to very large mass ratios or periods), such
that future observations with large scale optics should lead to accurate 
distance determinations. As emphasized in Section 4, we anticipate a coming time 
when several tests will better 
sift through high quality light curves and velocities from large optics. Comparison of 
photometric and spectroscopic mass ratios will then settle most of the otherwise unclear decisions
between detached and \emph{SD} assignments.

Although we expected that most binaries selected as \emph{SD} would be common 
normal Algols, the result is that only a small minority of our \emph{SD}'s have light curves
like those of common Algols. Indeed, inspection of the OGLE catalog reveals that systems 
with light-curves resembling those of classical Algols are virtually absent.
Only two of our \emph{SD} candidate binaries have solutions consistent 
with normal Algols. Basically the secondary stars have high 
surface brightnesses (\textit{i.e.} are hot compared to those of normal Algols). 
This outcome could be a strongly positive one, as \emph{SD}'s with bright 
secondaries will have stronger light curve solutions
(with information from \textit{two} deep eclipses) than ordinary Algols, while they
may retain all the \emph{SD} advantages mentioned in the Introduction. They also have
moderate luminosity ratios, and therefore greatly increased likelihood of their 
spectra being double-lined.

It could be that some of the 36 \emph{SD} binaries actually are \emph{OC}, as testing for
the \emph{OC} condition is beyond the scope of this paper and will be treated in future 
work. That outcome also could be advantageous, as \emph{OC} configurations can
have very well conditioned light curve solutions, perhaps even better than \emph{SD}'s.
\emph{OC} binaries are rare at the high luminosities detectable by OGLE (although very
common at much lower luminosity), yet discovery of just a few would be valuable help 
in finding the distance to the Magellanic Clouds.

 The next step is to obtain spectra and accurate multi-band light curves of the more 
promising systems so as to
confirm \emph{SD} assignments via spectral types and improved light curve parameters,
and also to observe radial velocities for absolute dimensions.

\acknowledgements{

\noindent  The authors are very grateful to Professor Bohdan Paczynski for 
suggestions that led to this project, and for reading the manuscript. 
We are indebted to the OGLE collaboration for making their data public domain, and hence 
projects like this possible. Finally we would like to thank the anonymous referee whose 
careful reading and comments significantly enhanced the papers clarity. This work was 
supported in part by NASA through a Hubble Fellowship grant from the Space Telescope 
Science Institute, which is operated by the Association of Universities for Research in
Astronomy, Inc., under NASA contract NAS 5-26555 (for J.S.B.W.), and in part by NSF 
grants AST-9819787 and AST-9820314 to Professor Paczynski. 

}

\newpage

\clearpage 

\begin{table*}
\begin{center}
\begin{scriptsize}
\begin{tabular}{ccccccc} 

Mode 2           &                 &             &    &              &                 &                   \\
Group 1          &                 &             &    &     Group 2  &                 &                   \\\cline{1-3}\cline{5-7}
subset 1         &subset 2         &subset 3     &    & subset 1     &subset 2         &subset 3           \\\cline{1-3}\cline{5-7}\cline{1-3}\cline{5-7}
  $e$            &       $i$       &    $T_1$    &    &   $e$        &       $i$       &   $T_1$           \\
  $\Omega_1$     &   $L_1$         &     $t_0$   &    &   $t_0$      &   $\Omega_2$    &  $\Omega_1$       \\
  $\Omega_2$     &                 &             &    &   $L_1$      &                 &                   \\\cline{1-3}\cline{5-7}\\\\
\end{tabular}
\vspace{5mm}

\begin{tabular}{ccccccc} 
Mode 5           &                 &             &    &              &                 &                   \\
Group 1          &                 &             &    &     Group 2  &                 &                   \\\cline{1-3}\cline{5-7}
subset 1         &subset 2         &subset 3     &    & subset 1     &subset 2         &subset 3           \\\cline{1-3}\cline{5-7}\cline{1-3}\cline{5-7}
  $e$            &       $i$       &    $T_1$    &    &   $e$        &       $i$       &   $T_1$           \\
  $\Omega_1$     &   $L_1$         &     $t_0$   &    &   $t_0$      &   $q$           &  $\Omega_1$       \\
  $q$            &                 &             &    &   $L_1$      &                 &                   \\\cline{1-3}\cline{5-7}\\
Group 3          &                 &             &    &     Group 4  &                 &                   \\\cline{1-3}\cline{5-7}
subset 1         &subset 2         &subset 3     &    & subset 1     &subset 2         &subset 3           \\\cline{1-3}\cline{5-7}\cline{1-3}\cline{5-7}
  $\Omega_1$     &       $i$       &    $T_1$    &    &   $t_0$        &       $i$       &   $T_1$           \\
  $q$            &   $L_1$         &     $t_0$   &    &   $L_1$      &   $q$           &  $\Omega_1$       \\\cline{1-3}\cline{5-7}\\\\
\end{tabular}
\end{scriptsize}

\caption{\label{t1} Table showing the groups of subsets used by dc in modes 2 (top) and 5 (bottom).}

\end{center}
\end{table*}


\begin{table*}
\begin{center}
\begin{scriptsize}
\begin{tabular}{lc} \hline

 parameter                            &  description \\\hline\hline

 $i$        (adjusted*)                &  binary orbit inclination (degrees).\\
 $T_1$      (adjusted*)                &  mean surface effective temp. (K) of star 1.\\
 $L_1$      (adjusted*)                &  luminosity for star 1.\\
 $t_0$       (adjusted)                &  zero point of orbital ephemeris.\\
 $e$          (adjusted)                &  binary orbit eccentricity.\\\\
 $\Omega_1$  (adjusted*)               &  potential of star 1.\\\\
 $\Omega_2$ (in mode 2: adjusted*)               &  potential of star 2.\\
 $\Omega_2$ (in mode 5: set by $q$)               &  potential of star 2.\\\\
 $q$  (in mode 2: set from initial guess)               &  mass ratio.\\
 $q$  (in mode 5: adjusted*)               &  mass ratio.\\\\

 $\omega$=  0.0  or $\pi$   (in mode 2: set from initial guess)   &  argument of periastron for star 1. \\
 $0<\omega<\pi$    (in mode 5: set from initial guess)              &  argument of periastron for star 1. \\\\

 $T_2$    =  10000.0                 &  mean surface effective temp. (K) of star 2.\\
 $L_2$      (set from initial guess)    &  luminosity for star 2.\\
 $P$     (from Udalski et al.)     &  period of binary orbit (days).\\
 $\lambda_I$=  0.9                   &  wavelength of light curve in microns.\\
 $x_1$      =  0.32                  &  linear limb darkening coefficient of star 1.\\
 $x_2$      =  0.32                  &  linear limb darkening coefficient of star 2.\\
 $y_1$      =  0.18                  &  non-linear limb darkening coefficient of star 1.\\
 $y_2$      =   0.18                 &  non-linear limb darkening coefficient of star 2.\\
 $l_3$      =  0.0                   &  third light.\\
 $f_1$       =  1.0                     &  ratio of axial rotation rate to mean orbital rate.\\
 $f_2$       =  1.0                     &  ratio of axial rotation rate to mean orbital rate.\\
 $g_1$      =  1.0                     &  exponent in gravity brightening (bolo. flux prop. to local gravity).\\
 $g_2$      =  1.0                     &  exponent in gravity brightening (bolo. flux prop. to local gravity).\\
 $A_1$     =  1.000                   &  bolometric albedo of star 1.\\
 $A_2$     =  1.000                   &  bolometric albedo of star 2.\\
 $\lambda$ =  $10^{-5}$              &  the Marquardt multiplier.\\\hline
\end{tabular}
\end{scriptsize}

\caption{\label{t2} Table of parameters with descriptions. The adjusted parameters are labeled as such
 (those for which convergence is required are also marked by *), and the values of fixed parameters are given.
 Note that g should not be confused with surface gravity.}

\end{center}
\end{table*}

\begin{table*}
\begin{center}
\begin{scriptsize}
\begin{tabular}{cc} \hline

 control integer                    &  description \\\hline\hline
 
 NREF     =  1                       &  number of reflections.\\
 MREF     =  1                       &  simple reflection treatment.\\
 LD       =  2                       &  logarithmic limb darkening law.\\
 JDPHS    =  1                       &  independent variable time.\\
 NOISE    =  1                       &  scatter scales with sqrt (light level).\\
 MODE     =  2 or 5                  &  mode of program operation.\\
 IPB      =  0                       &  for normal operation in mode 2.\\
 IFAT1    =  0                       &  for black body (star 1).\\
 IFAT2    =  0                       &  for black body (star 2).\\
 N1       =  30                      &  grid size for star 1.\\
 N2       =  30                      &  grid size for star 2.\\
 N1L       =  15                     &  coarse grid integers for star 1.\\
 N2L       =  15                     &  coarse grid integers for star 2.\\
 IFVC1     =  0                      &  no radial velocity curve for star 1. \\
 IFVC2     =  0                      &  no radial velocity curve for star 2.\\
 NLC       =  1                      &  number of light-curves.\\
 KDISK     =  0                      &  no scratch pad.\\
 ISYM      =  1                      &  use symmetrical derivatives.\\\hline

\end{tabular}
\end{scriptsize}

\caption{\label{t3} Table of control integers with descriptions (nomenclature from Wilson 1998).}

\end{center}
\end{table*}


\begin{table*}
\begin{tiny}
\begin{tabular}{ccccccccccccc} \hline
Field   &  Object &   $q$       & $\frac{R_1}{a}$  & $\frac{R_2}{a}$  &  $\frac{J_1}{J_2}$ &$i$ (degrees)& $e$          & $\frac{L_1}{L_2}$  & $F_e$      & Period  &  I    & V-I  \\\hline\hline
3 & 63551 & 0.412 $\pm$ 0.077 & 0.137 $\pm$ 0.010 & 0.282$\pm$0.014 & 2.599 & 84.1 $\pm$ 1.8 & 0.012 $\pm$ 0.015 & 0.54 &1.151& 2.9691 & 16.7090 & -0.0280 \\
3 & 189215 & 0.487 $\pm$ 0.040 & 0.151 $\pm$ 0.005 &0.298 $\pm$0.006 & 2.054 & 83.3 $\pm$ 0.8 & 0.000 $\pm$ 0.002 & 0.47 & 1.099& 2.6424 & 16.4560 & -0.0290 \\
3 & 213548 & 0.029 $\pm$ 0.006 & 0.251 $\pm$ 0.017 &0.128 $\pm$0.009 & 2.958 & 88.4 $\pm$ 6.3 & 0.013 $\pm$ 0.007 & 9.73 & 1.374& 4.0572 & 17.9280 & 0.3000 \\
4 & 53991 & 0.468 $\pm$ 0.071 & 0.130 $\pm$ 0.010 &0.294 $\pm$0.012 & 1.625 & 80.8 $\pm$ 1.2 & 0.002 $\pm$ 0.006 & 0.28 & 1.020& 4.1593 & 17.2130 & 0.1080 \\
4 & 91631$\dagger$ & 17.142 $\pm$ 1.912 & 0.080 $\pm$ 0.015 &0.565 $\pm$0.006 & 2.686 & 84.7 $\pm$ 3.6 & 0.000 $\pm$ 0.003 & 0.05 & 3.455& 4.5871 & 16.8760 & -0.0180 \\
4 & 163552 & 0.207 $\pm$ 0.010 & 0.383 $\pm$ 0.004 &0.234 $\pm$0.003 & 1.027 & 82.8 $\pm$ 0.7 & 0.004 $\pm$ 0.002 & 2.52 & 1.050& 1.5458 & 15.7350 & 0.0040 \\
5 & 190577$\dagger$ & 17.026 $\pm$ 2.233 & 0.082 $\pm$ 0.018 &0.561 $\pm$0.007 & 2.564 & 76.2 $\pm$ 3.3 & 0.013 $\pm$ 0.007 & 0.05 & 2.476& 3.4405 & 16.5630 & 0.0030 \\
5 & 266015 & 1.014 $\pm$ 0.051 & 0.183 $\pm$ 0.003 &0.357 $\pm$0.004 & 1.685 & 80.8 $\pm$ 0.4 & 0.000 $\pm$ 0.002 & 0.39 & 1.037 & 1.8089 & 15.9020 & -0.0320 \\
6 & 163076 & 0.132 $\pm$ 0.007 & 0.378 $\pm$ 0.004 &0.206 $\pm$0.003 & 0.998 & 81.5 $\pm$ 0.7 & 0.003 $\pm$ 0.003 & 3.04 & 1.058 & 1.6684 & 16.7660 & -0.0660 \\
6 & 242137 & 0.108 $\pm$ 0.009 & 0.352 $\pm$ 0.007 &0.192 $\pm$0.005 & 1.162 & 81.5 $\pm$ 1.0 & 0.013 $\pm$ 0.006 & 3.47 & 1.032 & 2.4572 & 17.2040 & - \\
6 & 306609 & 1.061 $\pm$ 0.089 & 0.167 $\pm$ 0.004 &0.361 $\pm$0.007 & 1.165 & 90.0 $\pm$ 14.9 & 0.002 $\pm$ 0.002 & 0.22 & 1.580 & 2.1532 & 17.0150 & 0.0370 \\
9 & 59110 $*$& 1.963 $\pm$ 0.276 & 0.099 $\pm$ 0.009 &0.413 $\pm$0.012 & 2.320 & 89.7 $\pm$ 5.9 & 0.001 $\pm$ 0.002 & 0.12 & 2.564 & 16.3409 & 17.7390 & 0.9050 \\
10 & 3097 & 1.374 $\pm$ 0.128 & 0.184 $\pm$ 0.005 &0.381 $\pm$0.008 & 1.077 & 85.1 $\pm$ 1.4 & 0.005 $\pm$ 0.002 & 0.23 & 1.303 & 2.1284 & 16.3150 & -0.1410 \\\hline

1 & 27259 & 0.517 $\pm$ 0.088 & 0.147 $\pm$ 0.014 &0.296 $\pm$0.013 & 2.366 & 81.0 $\pm$ 1.3 & 0.028 $\pm$ 0.012 & 0.52 & 0.976 & 2.6569 & 17.7680 & -0.0120 \\
1 & 32309 & 0.739 $\pm$ 0.097 & 0.191 $\pm$ 0.018 &0.328 $\pm$0.011 & 2.311 & 73.6 $\pm$ 0.4 & 0.011 $\pm$ 0.005 & 0.70 & 0.620 & 3.8188 & 16.4290 & -0.1200 \\
3 & 28139$*$ & 0.146 $\pm$ 0.034 & 0.096 $\pm$ 0.011 &0.211 $\pm$0.014 & 1.499 & 81.3 $\pm$ 0.6 & 0.007 $\pm$ 0.004 & 0.27 & 0.812 & 21.2494 & 17.0440 & 0.3880 \\
4 & 91721 & 1.143 $\pm$ 0.110 & 0.214 $\pm$ 0.016 &0.359 $\pm$0.008 & 2.686 & 74.9 $\pm$ 0.7 & 0.032 $\pm$ 0.013 & 0.87 & 0.731 & 2.3136 & 17.7780 & 0.0250 \\
4 & 117831 & 0.157 $\pm$ 0.054 & 0.357 $\pm$ 0.032 &0.209 $\pm$0.021 & 1.010 & 77.9 $\pm$ 2.1 & 0.049 $\pm$ 0.019 & 2.69 & 0.852 & 1.1645 & 17.7790 & -0.0780 \\
4 & 160001$*$ & 0.051 $\pm$ 0.014 & 0.148 $\pm$ 0.010 &0.155 $\pm$0.013 & 0.996 & 81.3 $\pm$ 0.6 & 0.002 $\pm$ 0.004 & 0.78 & 0.513 & 23.4661 & 17.6730 & 0.6280 \\
5 & 239994 & 0.996 $\pm$ 0.102 & 0.210 $\pm$ 0.018 &0.349 $\pm$0.008 & 2.719 & 75.5 $\pm$ 0.5 & 0.024 $\pm$ 0.009 & 0.88 & 0.737 & 4.7274 & 16.8870 & 0.0530 \\
5 & 271572 & 0.553 $\pm$ 0.077 & 0.197 $\pm$ 0.021 &0.303 $\pm$0.011 & 2.468 & 80.1 $\pm$ 0.6 & 0.019 $\pm$ 0.008 & 0.92 & 0.834 & 2.2733 & 17.6940 & 0.0590 \\
5 & 316901 & 0.912 $\pm$ 0.120 & 0.202 $\pm$ 0.018 &0.345 $\pm$0.011 & 2.252 & 79.3 $\pm$ 0.7 & 0.011 $\pm$ 0.008 & 0.69 & 0.894 & 1.5304 & 17.3330 & -0.0060 \\
6 & 277060 & 0.848 $\pm$ 0.142 & 0.203 $\pm$ 0.021 &0.335 $\pm$0.014 & 2.671 & 78.2 $\pm$ 0.7 & 0.030 $\pm$ 0.014 & 0.88 & 0.821 & 3.1612 & 16.4920 & - \\
7 & 66175 & 1.060 $\pm$ 0.193 & 0.195 $\pm$ 0.027 & 0.358$\pm$0.015 & 1.303 & 76.8 $\pm$ 1.5 & 0.012 $\pm$ 0.008 & 0.34 & 0.833 & 3.6264 & 14.3820 & -0.0140 \\
7 & 110260 & 0.744 $\pm$ 0.111 & 0.125 $\pm$ 0.015 &0.311 $\pm$0.012 & 2.497 & 72.5 $\pm$ 0.5 & 0.001 $\pm$ 0.003 & 0.31 & 0.620 & 3.2881 & 16.5900 & -0.0500 \\
7 & 228350 & 0.560 $\pm$ 0.120 & 0.204 $\pm$ 0.031 &0.302 $\pm$0.017 & 2.517 & 77.3 $\pm$ 0.9 & 0.027 $\pm$ 0.015 & 1.02 & 0.703 & 5.0240 & 17.0900 & 0.0810 \\
7 & 243913 & 1.447 $\pm$ 0.134 & 0.208 $\pm$ 0.019 &0.386 $\pm$0.008 & 1.365 & 76.5 $\pm$ 1.3 & 0.004 $\pm$ 0.006 & 0.36 & 0.869 & 2.6316 & 14.8400 & -0.0960 \\
8 & 87175 & 0.968 $\pm$ 0.076 & 0.218 $\pm$ 0.009 &0.354 $\pm$0.006 & 1.005 & 81.8 $\pm$ 1.1 & 0.001 $\pm$ 0.001 & 0.34 & 0.985 & 1.1022 & 16.2200 & -0.1100 \\
8 & 110280 & 0.655 $\pm$ 0.053 & 0.210 $\pm$ 0.010 &0.320 $\pm$0.006 & 2.048 & 81.3 $\pm$ 0.4 & 0.007 $\pm$ 0.007 & 0.79 & 0.899 & 3.6153 & 16.0050 & -0.0770 \\
8 & 170230 & 0.384 $\pm$ 0.074 & 0.220 $\pm$ 0.021 &0.272 $\pm$0.014 & 2.512 & 79.8 $\pm$ 0.7 & 0.035 $\pm$ 0.013 & 1.46 & 0.716 & 5.0538 & 17.2660 & 0.1880 \\
9 & 41849 & 0.867 $\pm$ 0.054 & 0.164 $\pm$ 0.008 &0.342 $\pm$0.005 & 2.387 & 76.1 $\pm$ 0.4 & 0.010 $\pm$ 0.005 & 0.49 & 0.812 & 3.1590 & 16.4650 & -0.1010 \\
9 & 94700 & 0.958 $\pm$ 0.206 & 0.231 $\pm$ 0.030 &0.348 $\pm$0.018 & 2.677 & 83.1 $\pm$ 2.0 & 0.017 $\pm$ 0.016 & 1.07 & 0.992 & 2.1934 & 18.3980 & 0.0610 \\
9 & 163573 & 0.026 $\pm$ 0.007 & 0.402 $\pm$ 0.018 &0.118 $\pm$0.011 & 1.831 & 72.0 $\pm$ 2.5 & 0.076 $\pm$ 0.032 & 18.8 & 0.895 & 1.1670 & 14.6640 & -0.2180 \\
10 & 47428$*$ & 0.138 $\pm$ 0.032 & 0.148 $\pm$ 0.021 &0.207 $\pm$0.014 & 1.539 & 85.5 $\pm$ 1.1 & 0.011 $\pm$ 0.006 & 0.68 & 0.936 & 23.4192 & 18.4770 & 1.0010 \\
10 & 120906$*$ & 1.129 $\pm$ 0.076 & 0.145 $\pm$ 0.006 &0.366 $\pm$0.006 & 1.592 & 76.6 $\pm$ 0.8 & 0.002 $\pm$ 0.003 & 0.22 & 0.962 & 89.0876 & 16.0910 & 1.3120 \\
11 & 58186$*$ & 0.045 $\pm$ 0.015 & 0.117 $\pm$ 0.018 &0.146 $\pm$0.016 & 1.835 & 81.0 $\pm$ 1.0 & 0.022 $\pm$ 0.023 & 1.01 & 0.459 & 10.5115 & 18.3500 & 0.1480 \\\hline

\end{tabular}
\end{tiny}

\caption{\label{t4} Table of parameters for \emph{SD} eclipsing binaries.
Those having complete eclipses are shown in the upper section, and 
those having partial eclipses in the lower section of the table. The periods (in days),
 I-magnitudes and colors are taken from Udalski et al. (1998). The binaries 
marked by $\dagger$'s are doubtful due to very large mass ratios, and those
marked by an $*$ are doubtful due to long periods. }

\end{table*}

\end{document}